\documentclass[10pt,
amsmath,amssymb,
aps,prd,nofootinbib,eqsecnum,a4paper]{revtex4}
\usepackage{graphicx,epsf}
\usepackage{float}
\usepackage{wrapfig}
\usepackage{xcolor}
\usepackage[colorlinks = true,
            linkcolor = blue,
            urlcolor  = blue,
            citecolor = blue,
            anchorcolor = blue]{hyperref}
\def\be{\begin{equation}}
\def\ee{\end{equation}}
\def\bea{\begin{eqnarray}}
\def\eea{\end{eqnarray}}

\usepackage{framed}

\usepackage{float}

\newcommand{\CC}{C\nolinebreak\hspace{-.05em}\raisebox{.4ex}{\tiny\bf +}\nolinebreak\hspace{-.10em}\raisebox{.4ex}{\tiny\bf +}}
\def\CC{{C\nolinebreak[4]\hspace{-.05em}\raisebox{.4ex}{\tiny\bf ++}}}
\def\S{ }
\begin{document}

\title{PyTransport: A Python package for the calculation of inflationary correlation functions}
\author{David J. Mulryne and John W. Ronayne}
\affiliation{Astronomy Unit, School of Physics and Astronomy,
Queen Mary University of London, Mile End Road, London, E1 4NS, UK}
\email[]{d.mulryne@qmul.ac.uk} 
\email[]{j.ronayne@qmul.ac.uk}

\date{\today}

\begin{abstract}

\noindent PyTransport constitutes a straightforward code written in \CC \S  together 
with Python scripts which automatically edit, compile and run the \CC \S code as a 
Python module. It has been written for Unix-like systems (OS X and Linux).
Primarily the module employs the transport approach to inflationary cosmology to calculate 
the tree-level power-spectrum and bispectrum of user specified models of multi-field inflation, 
accounting for all sub and super-horizon effects.
The transport method we utilise means 
only coupled differential equations need to be solved, and the implementation presented here 
combines the speed of \CC \S  with the functionality and convenience of Python. This document details the code and illustrates how to use it with a worked example. It has been updated to 
be a companion to the second version of the code, 
PyTransport\,2.0, which includes functionality to deal with models of inflation with a curved 
field space metric.

\end{abstract}

\maketitle




\begin{figure}[H]
\centering
\includegraphics[width=11cm]{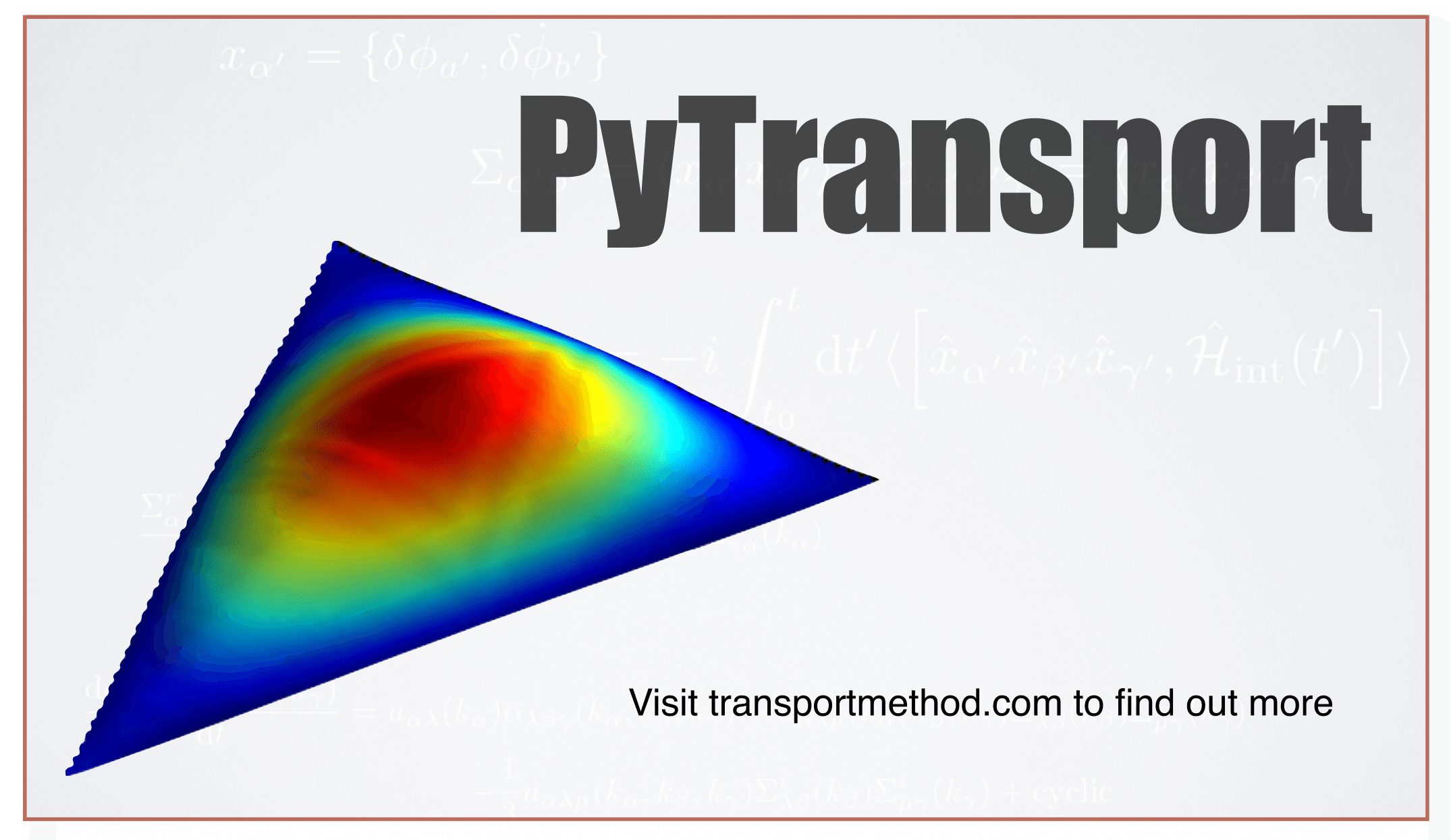}
\end{figure}

\section{Introduction}

\begin{framed}
{\bf \noindent PyTransport is distributed under a GNU GPL licence. The most recent version can be obtained by visiting \href{https://transportmethod.com}{transportmethod.com}. If you use PyTransport you are kindly asked to cite Ref.~\cite{Dias:2016rjq}  as well as the archive version of this user guide in any resulting works. }
\end{framed}

The main purpose of this document is to teach those interested how to use, and if so desired adapt, 
the PyTransport package. It has now been updated  to 
be a companion to the second version of the code, 
PyTransport\,2.0, which includes functionality to deal with models of inflation with a curved 
field space metric. We have had to make some minimal syntax changes in this second version 
in order to support new functionality, as discussed below. Users of the original package 
will unfortunately not be able to switch to the new one without amending their scripts. The original user guide 
can still be found as the arXiv version~1 of this document.
The philosophy behind the implementation is simplicity and ease of use. 
Python was selected as the 
language though which to interact with the code because it enables rapid scripting and provides a flexible 
and powerful platform. 
In particular, it has many readily available tools and packages for analysis and visualisation, and for tasks such as parallelisation  (using for example Mpi4Py).  
As an interpreted language, however, Python can be slow for some tasks. This is circumvented here by 
using \CC \S  code, 
which is compiled into a Python module, to perform numerically intensive tasks with the result that the speed 
of the package is nearly indistinguishable from pure \CC. The \CC \S  code itself is kept as simple and 
clean as possible and can therefore easily be edited if required. PyTransport has been developed on 
OS X 
using Python 2.7. We have also performed limited testing on Linux systems, and 
attempted to ensure compatibility with versions of Python 3. 
It can also be adapted to Windows systems, but this functionality has not yet been incorporated into 
the released package\footnote{We thank Sean Butchers for work related to installing PyTransport on a 
Windows machine.}

The code is intended to be a reusable resource for inflationary cosmology. It enables users to quickly create a 
complied Python module(s) for any given model(s) of multi-field inflation. 
The primary function of the complied module is to calculate the power-spectrum and bi-spectrum of inflationary 
perturbations produced by multi-field inflation. To this end,
the module contains a number 
functions that can be called from Python and that perform tasks such as calculating the background evolution 
of the cosmology, as well as the evolution of the two and three point functions. We also provide a number of further functions written in 
Python that perform common tasks such as calculating the power spectrum or bispectrum over a range of scales by utilising the 
compiled module.
The true power of the approach, however, is that users can rapidly write their own scripts, or adapt ours, to suit their own needs. 

The transport approach to inflationary perturbation theory that the code employs 
can be seen as the differential version of the integral expressions of the In-In formalism. It 
is helpful numerically because it provides a set of ordinary differential equations for the correlation functions  
of inflationary perturbations. The code solves these equations from deep inside the horizon until some desired time 
after horizon crossing using a standard variable step size ordinary differential equation (ODE) 
routine with error control. Such off the shelf 
routines are extremely well tested, and provide
an easy way to change the required accuracy. This is helpful in order to check convergence of the numerical 
solutions, or to respond to needs of models with very fine features. 
Details of the transport method itself that the code is based on can be found in the recent papers \cite{Dias:2016rjq} 
and \cite{xxx2}, the second of which updates the method to allow
for the analysis of models with a curved field space metric. We 
highly recommend reading this guide in combination with those papers.

In this guide, we first we give some brief background and motivation for the code, much more can be found in Refs.~\cite{Dias:2016rjq,xxx2}, 
before giving an overview of its structure and how 
it can be set up. In the appendices we give some more detail about the structure of the underlying \CC \S code,  
give full details of all the functions the complied module provides, and all the functions provided by 
Python scripts which accomplish common tasks. The best way to learn how to use the package, however, is 
by example. We present an extended example below spread between the ``Getting going"  and ``Examples" sections, 
complete with screen shots of the code in use. 
Other examples that come with the distribution are discussed in the ``Examples" section.
Throughout, familiarity with Python and to some extent \CC \S  is assumed, though in reality users 
can just probably get a long way by looking at the examples and modifying to their needs.

Finally, we would also like to refer readers to the complementary package developed 
in tandem with the work in Ref.~\cite{Dias:2016rjq} and with PyTransport: CppTransport \cite{Seery:2016lko}. This is a platform for inflationary 
cosmology developed fully in \CC and recently also 
updated to deal with curved field space metrics \cite{seeryNew}. In comparison with PyTransport it  has 
more external dependancies (in the sense that the dependancies of PyTransport are mainly 
Python modules), but provides more sophisticated parallelisation and data management capabilities. 
In limited testing it is also found to be marginally faster.
For users with modest aims in terms of CPU hours and data generation, however, it is likely to have a higher 
overhead 
in getting started, but may well be beneficial for intensive users. PyTransport 
is intended to be more lightweight with users encouraged to utilise the power of Python in combination 
with PyTransport to achieve their specific aims and data management needs.

\section{Background} 

Calculations of the correlation functions of perturbations produced by inflation 
are now extremely mature.
In the single field context, the In-In formalism is routinely used to calculate the equal time correlation 
functions of the 
curvature perturbation, $\zeta$, as wavelengths cross the cosmological horizon 
\cite{Maldacena:2002vr, Seery:2005wm, Chen:2006nt,Elliston:2012ab} where they become 
constant \cite{Rigopoulos:2003ak,Lyth:2004gb}. 
For many models this calculation can be accurately performed 
analytically, while for others, such as models with features, 
a numerical implementation is required \cite{Chen:2006xjb,Chen:2008wn,Hazra:2012yn,Funakoshi:2012ms}. 
If additional fields are added, the problem becomes even more complex. $\zeta$ is no longer 
necessarily conserved after horizon crossing, and the evolution of all 
isocurvature modes needs to be accounted for --    
all the way from the initial vacuum state until such time as the system becomes adiabatic, or until the time at which we wish to know the statistics  (see for example Ref.~\cite{Elliston:2011dr} 
for a discussion of adiabaticity). While analytic progress can be made in some circumstances 
using the In-In formalism and/or so called ``super-horizon" techniques such as the $\delta N$ formalism, 
in general for multiple field models numerical techniques become even more important. 

The code documented here accounts for all tree-level effects present in multi-field inflation. This includes  
the super-horizon evolution of $\zeta$, which 
can occur in models with multiple light  
fields, as well as the effect of features in the multidimensional potential, and the effect of 
quasi-light or heavy fields orthogonal to the inflaton 
(which are important if the inflationary trajectory is not straight).
As discussed above the code utilises the transport approach to inflationary correlation functions 
\cite{Mulryne:2009kh,Mulryne:2010rp,Dias:2011xy,Anderson:2012em,Seery:2012vj, Mulryne:2013uka}. This approach 
can be viewed as a differential version of the integral expressions of the In-In formalism, 
and evolves correlations of inflationary perturbations from their vacuum state on 
sub-horizon scales until we wish to evaluate the statistics. 
We note for clarity that in its original form the ``moment transport'' method was restricted 
to super-horizon scales, but it was later shown how it could be extended to include sub-horizon scales 
in Ref.~\cite{Mulryne:2013uka}. A recent paper studies this extension further and develops it into a working 
algorithm \cite{Dias:2016rjq} with many additional details provided. The present document details 
the code PyTransport which is discussed in that paper.

At the background level an inflationary cosmology is completely determined by the evolution of the 
fields, $\phi_i$, and their associated velocities (the rate at which the fields change with cosmic time), $\dot \phi_i$,
as a function of the number of e-folds (the logarithm of the scale factor, $N=\ln(a)$) which occurs.
At the perturbed level, the key objects are the 
correlations of the perturbations in these fields, 
and correlations of other perturbative quantities.  Here we have used the label $i$ 
to run over the number of fields present. 
The code numerically solves the equations of motion for the background fields, and 
the equations of motion for the evolution of correlations of the field and field velocity perturbations 
defined on flat hyper-surfaces. Ultimately the quantities 
probably of most interest for observations are the statistics of the curvature perturbation 
$\zeta$ -- in particular the power spectrum and bispectrum -- which the code calculates from the 
field space correlations. 
Defining the array ${\mathbf X} = \{{ Q^I}, { P^J} \}$ of the covariant field space perturbation and its momentum\footnote{For canonical models where the field space metric is Euclidean $Q^I$ is simply the field space perturbation $\delta \phi^I$, see Refs.~(\cite{Gong:2011uw,Elliston:2012ab}) for its definition in the case of a non-Euclidean metric.}, where 
the components are labelled $X^a$ and $a$ now runs over the total number of fields and field velocities, 
we recall the following definitions for later clarity:
\begin{eqnarray}
\langle \zeta(\mathbf{k_1}) \zeta(\mathbf{k_2})  \rangle &=& (2\pi)^3 \delta(\mathbf{k}_1 + \mathbf{k}_2 )
        P_\zeta(k_1) \,, \\
\langle X^a(\mathbf{k_1}) X^b(\mathbf{k_2})  \rangle &=& (2\pi)^3 \delta(\mathbf{k}_1 + \mathbf{k}_2 )
        \Sigma^{ab}(k_1) \,, \\
\langle \zeta(\mathbf{k_1}) \zeta(\mathbf{k_2}) \zeta(\mathbf{k_3}) \rangle &=& (2\pi)^3 \delta(\mathbf{k}_1 + \mathbf{k}_2 + \mathbf{k}_3)
        B_\zeta(k_1, k_2, k_3) \,, \\
\langle X^a(\mathbf{k_1}) X^b(\mathbf{k_2}) X^c(\mathbf{k_3}) \rangle &=& (2\pi)^3 \delta(\mathbf{k}_1 + \mathbf{k}_2 + \mathbf{k}_3)
        B^{abc}(k_1, k_2, k_3)\,, \\
        \frac{6}{5}f_{\rm NL}(k_1,k_2,k_3) &=& B_\zeta(k_1,k_2,k_3)/ \left( P_\zeta(k_1)P_\zeta(k_2) + P_\zeta(k_2)P_\zeta(k_3) + P_\zeta(k_1)P_\zeta(k_3) \right ) \, , 
\end{eqnarray}
where $P_\zeta$ and $B_\zeta$ are power spectrum and bispectrum of $\zeta$ respectively, 
and $\Sigma$ and $B$ are 
the equivalent functions for the correlations and cross correlations in field space, and $f_{\rm NL}$ is 
the reduced bispectrum.
$\Sigma$ and $B$ together with the background 
values of the fields and field velocities are the objects directly evolved by the code using the equations 
detailed in section~5 of Ref.~\cite{Dias:2016rjq} with 
initial conditions 
detailed in section~6 of Ref.~\cite{Dias:2016rjq}\footnote{Internally the code internally rescales 
the field velocity perturbation $P^I$ for 
performance reasons, as well as internally rescaling the wavenumbers involved, but the rescaling is reversed before 
outputting results -- this is discussed further in appendix 1.}. As 
discussed in that paper, $B$ and $\Sigma$ can then 
readily be converted to give $P_\zeta$ and $B_\zeta$ through the use of the ``$N$" tensors with components $N_a$ and $N_{ab}$, 
described in section~7 of Ref~\cite{Dias:2016rjq} and updated to the case of a curved field space metric in Ref.~\cite{xxx2} (see also \cite{Dias:2014msa}). 
The equations of motions for the correlations 
are given in Eqs.~(5.5) and (5.16) of  Ref.~\cite{Dias:2016rjq}, the initial conditions in Eqs.~(6.2) and (6.7),
and the conversion to $\zeta$ in 
Eq.~(7.4).

It is worth briefly commenting on how our code compares with existing ones. A number of publicly available 
codes exist to calculate the power spectrum from canonical multi-field inflation. 
For example Pyflation \cite{Huston:2011fr} and 
MultiModeCode \cite{Price:2014xpa} employ a method originally used by Salopek and Bond \cite{Salopek:1988qh}
in which the mode functions of the QFT of inflationary perturbations are evolved. These codes are 
in written Python\footnote{Pyflation also makes sure of C code for speed through the use of Cython.} and Fortran respectively.
Moreover, a Mathematica code which implements the transport method for curved field space 
metric models, but is restricted to the power-spectrum, is also available
\cite{Dias:2015rca}. At the level of the three-point function a number of authors 
have undertaken numerical work directly utilising 
the In-In formalism, for example in Refs.~\cite{Chen:2006xjb,Chen:2008wn, Hazra:2012yn,Funakoshi:2012ms,Horner:2013sea}. 
The only publicly released code we are aware of, however, is BINGO \cite{Hazra:2012yn} which is restricted to single field models. 
No general multi-field codes have been undertaken until now.

A further advantage of PyTransport (and CppTransport) over previous codes is that it leaves
little for the user to calculate analytically. 
It needs the user only 
to provide the inflationary potential. Then all the derivatives are automatically calculated using 
symbolic Python (SymPy) and written automatically into the \CC \S  code which is then compiled. 
Compared to Fortran or pure \CC \S  
implementations PyTransport 
has the advantage of easy access to the extensive and easy to use modules available to 
Python, and compared to a pure Python or Mathematica implementation we have the advantage of speed.

\section{ Code overview} 

\noindent  The code structure should become familiar though the extended example we provide, but here we give a brief summary. 

The code is distributed in a folder called {\it PyTransportDist/}\footnote{If downloaded from GitHub, it will instead come in folder named PyTransport and labeled with the branch of the code.}, which also contains a copy of this document (possibly updated compared with the arXiv version) in the 
{\it PyTransportDist/docs/} folder. The base code for PyTransport is written in \CC \S  and has a simple object orientated structure. This code can be found 
in the {\it PyTransportDist/PyTransport/CppTrans} folder and we provide a few more details about its structure 
and functionality in  appendix 1. 
The \CC \S code is deliberately as simple as possible to ensure transparency 
and adaptability. The idea of the PyTransport package as a 
whole is that after a potential and a field space metric (if the metric is non-Euclidean) are provided by the user 
the \CC \S  code is automatically edited and complied into a Python module by supporting 
Python functions (called from the {\it PyTransportDist/PyTransport/PyTransSetup.py} file which is described in full 
in appendix~2), 
meaning a lot of work is done for the user.  
The end result is a Python module consisting of a set of Python functions for a specific inflationary model, called the 
PyTrans*** module. 
The functions of this module provide key routines for inflationary cosmology (including calculating the evolution of the two and 
three point correlations). The asterisks, ***, indicate we can label the module with a tag telling us what model it 
corresponds to, and we can 
therefore install multiple modules if we want to work with many models simultaneously. The 
key functions available to these modules 
are defined in the file {\it PyTransportDist/PyTransport/PyTrans/PyTrans.cpp} (which is a \CC \S  file defining the
 Python module), these functions are detailed in appendix~3.
The scripts  that edit the \CC \S code and compile the module are discussed further below in the setup section, 
and by default they place the compiled module in the local folder 
{\it PyTransportDist/PyTransport/PyTrans/lib/python/} to avoid access issues if, for example, you do not have root privileges. 
Other useful Python functions that perform common tasks, such as producing a power spectrum by looping 
over calls to the compiled module, can be 
found in {\it PyTransportDist/PyTransport/PyTransScripts.py}, and we describe them below, and in full in detail in appendix~4. 
Python 
treats functions written in Python inside a file, such as {\it PyTransScripts.py} and {\it PyTransSetup.py}, 
in the same way as a compiled module. 
So there are effectively {\bf three modules within PyTransport}, 
one to setup a compiled module for the potential we want to study ({\bf PyTransSetup}), the compiled module itself ({\bf PyTrans***}) 
(or multiple complied modules labeled with different tags) and a 
module with various functions automating common tasks that 
use the functions of the compiled module ({\bf PyTransScripts}). 
Also in the {\it PyTransportDist/} folder is an example folder {\it PyTransportDist/Examples} 
containing the examples discussed below. 
There are no dependancies external to the folders provided except for a working Python 
installation (with appropriate packages downloaded), and a \CC \S compiler -- this is deliberate to make 
the code as easy as possible to use.
An MPI installation such as openMPI is also needed if the module is required to be 
used across multiple cores.

We note that all the \CC \S  code is written by the transport team except for an included Runge-Kutta-Fehlberg (rkf45)
integrator routine written by John Burkardt and distributed under a GNU LGPL license detailed \href{https://people.sc.fsu.edu/~jburkardt/f_src/rkf45/rkf45.html}{here}\footnote{${\rm https\hspace{-.1cm}:\hspace{-.1cm}//people.sc.fsu.edu/\hspace{-.1cm}\sim \hspace{-.1cm} jburkardt/f\_src/rkf45/rkf45.html}$}.  We choose this lightweight integrator over other options, such as using integrators included with 
the BOOST library, in order that it could easily be included with the distribution with no external dependancies being introduced. 
In our (limited) testing it functions well for all the models we have looked at.
There are no dependancies external to the folders provided except for a working Python 
installation (with appropriate packages downloaded), and a \CC \S compiler -- this is deliberate to make 
using the code as easy as possible to use.
An MPI installation such as openMPI is also needed if the module is required to be 
used across multiple cores. 

\section{Setup}
\label{Setup}

\subsection {Prerequisites} 

\noindent  So what is needed? The idea is as little as possible beyond Python:

\vspace{0.2cm}
\noindent {\bf Python:}  A working Python installation is needed, in development we used Python 2.7 which we recommend, 
but have subsequently attempted to ensure compatibility with versions of Python 3. 
For convenience we recommend downloading a complete Python distribution, for  
  example Enthought Canopy or Continuum Anaconda\footnote{One user has reported issues in getting mpi4py working with 
  their Anaconda installation on a mac, the author uses Canopy and has experienced no problems.}, which 
  come with all the core packages used by the code as well as 
  interactive development environments.
  Python packages currently used by PyTransport or by provided examples include Numpy, Matplotlib, SciPy, 
  Gravipy (needed only for 
  models with an non-trivial field space metric), SymPy, Distutils, Math and Sys, 
  as standard, and Mpi4Py and Mayavi are used 
  for MPI and 3D bispsectra plots respectively. Of these only Mpi4Py and Mayavi may 
  need to be downloaded separately from the distributions mentioned. One way to install a package 
  such as Mpi4Py is to type 
 ``pip install Mpi4Py" in the terminal. If using Canopy, Anaconda or similar, they come with their 
 own package managers  which are an even easier way to install packages. 
 There are many 
 easily found resources on the internet to help with such installations if there is a system 
 specific snag. Note that you should not attempt to install Mpi4Py without installing MPI first (which we deal with next). We also note that although apparently possible we have not easily been able to install Mayavi with Python 3.5, and recommend searching for online resources to help with this.

\vspace{0.2cm}
\noindent {\bf MPI:}
As computing the bispectrum can be computationally expensive, distributed computing can be helpful 
(even if only across the multiple cores of modern PCs). 
In some of the scripts in the 
{\it PyTransScripts} module,  
we use the Mpi4Py module to implement this. Mpi4Py needs a working MPI installation such as 
openMPI installed on your 
computer. Note that Mpi4py or openMPI are not needed for 
PyTransport in general, and if you do not have these installed you simply  
cannot run the scripts that use MPI, but can run the code in serial instead. A nice guide to installing openMPI is 
at this \href{https://wiki.helsinki.fi/display/HUGG/Open+MPI+install+on+Mac+OS+X}{link}\footnote{https://wiki.helsinki.fi/display/HUGG/Open+MPI+install+on+Mac+OS+X}.

\vspace{0.2cm}
\noindent {\bf \CC \S  complier:}
Python needs to be able to find a \CC \S  complier in order to compile the PyTrans module(s). This is bundled with most 
Linux distributions. If not present on a Mac system, downloading Xcode from the app store is the easiest way to 
install one (or Xcode command line tools can be downloaded separately, as explained \href{http://railsapps.github.io/xcode-command-line-tools.html}{here}\footnote{http://railsapps.github.io/xcode-command-line-tools.html}).

\subsection{Getting going}

\noindent  Once you have Python running and a \CC \S compiler, to get started take the {\it PyTransport/} folder from the 
{\it PyTransportDist/}\footnote{Or, If downloaded from GitHub, the code comes in a folder named PyTransport and labeled with the branch of the code, in which the {\it PyTransport/} folder can be found. For example {\it PyTransport-master/} .} folder and place it 
anywhere convenient in your computer's file system. It is essential that you don't change 
the structure of the sub-directories within {\it PyTransport/} , but you can place this folder wherever you want.  
That's more or less all you have to do.
You can do this by copying the entire folder {\it PyTransportDist/} (which 
also contains examples and this guide) to a convenient location, but equally well you could run examples from 
anywhere else on your computer. In each example, we will see one needs to add the path of  
the {\it PyTransport/} folder to the paths which Python includes when looking for code, so that Python can 
find the setup file 
{\it PyTransSetup.py} (or this could be done permanently). Now you can get started, no other installation is required which is not handled for you by provided 
Python scripts.

Lets say you want to analyse a canonical inflationary model (we will give an 
example of a model with a curved field space metric in Section\,\ref{NCexample})
defined by the potential \be
\label{dubquadpot}
V=\frac{1}{2} m^2_{\phi} \phi^2 + \frac{1}{2} m^2_{\chi} \chi^2\,.
\ee
The first step is to create (by compiling the \CC \S  code) a Python module for this potential. This is achieved by writing 
the potential in a Python file in SymPy (symbolic Python notation) and calling the 
appropriate functions from the { PyTransSetup module}. First define two SymPy 
arrays one for the fields and the other for parameters which 
define the model (the parameters you might wish to change the value of). 
These have length nF and nP respectively. 
Then define the symbolic potential $V$ using these arrays. The potential must be written into the 
\CC \S  code by calling the function {\it potential(V,nF,nP)}  which is in the file {\it PyTransSetup.py}.  If we 
wanted to work with a non-Euclidean field space metric it would also be specified at this stage, but if one is not specified explicitly 
the code assumes the model is canonical.
If working with a version of Python 2,  the next step is to call the function {\it compileName(``Quad")} (where ``Quad'' can be replaced by any name the user likes (the *** from above)). 
If using Python 3, users must use the function {\it compileName3(``Quad")}
to achieve the same thing. 
Multiple modules for different potentials with different names can therefore be created and 
used simultaneously. In the previous version of the code the tolerances used by the numerical 
integrator had to be specified at this setup stage, but these can now be set 
at the point we want to calculate a given evolution (and hence more easily adjusted). 
Below is a screen shot of the procedure just described with copious comments which should 
make the procedure clear:

\begin{figure}[H]
\centering
\includegraphics[width=18.0cm,height=7cm]{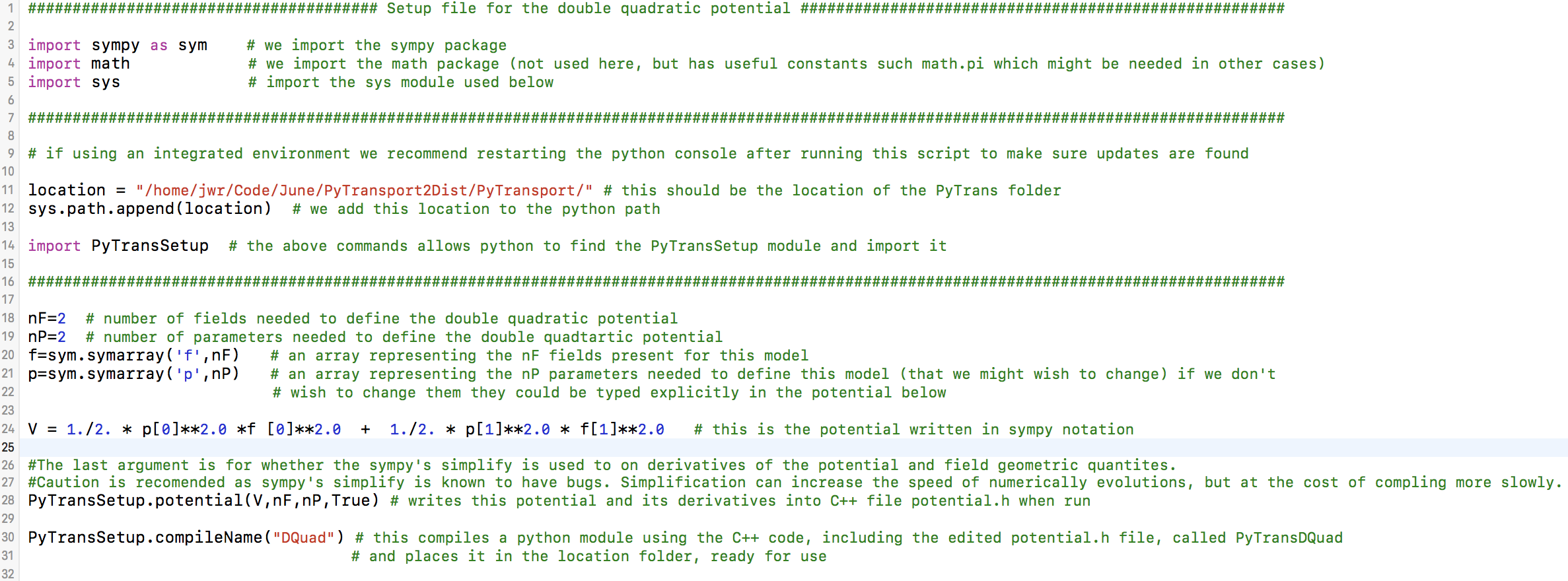}
\end{figure}

\noindent This example is contained in the {\it Examples/DoubleQuad/} folder which accompanies the code, with this script in the file 
{\it DQuadSetup.py} file. In this script 
we have used the two functions from the PyTransSetup module mentioned above. 
Appendix 2 contains a summary of all the functions 
available in the setup module.

The complied Python module can now be used. To do so we need to point Python to the path of the new module (and 
the scripts module if we wish to call the provided Python scripts). This can be done automatically 
by calling the function in the setup module {\it pathSet()}. We recommend using the {\it PyTransQuad} 
module in a separate file from the one used to set it up, and if working in an integrated development environment to restart the Python kernel (this is to ensure the most recent version is always imported). Below is a screen shot of the start of a file in which we use the module we set up in the previous paragraph. In the screen shot we first calculate the value of the potential and the first derivative of the potential 
for a particular choice of field values and parameters. Then we use 
these to set up an array containing field values and the associated fields velocity (using the slow roll
equation):

\begin{figure}[H]
\centering
\includegraphics[width=18cm, height=11.2cm]{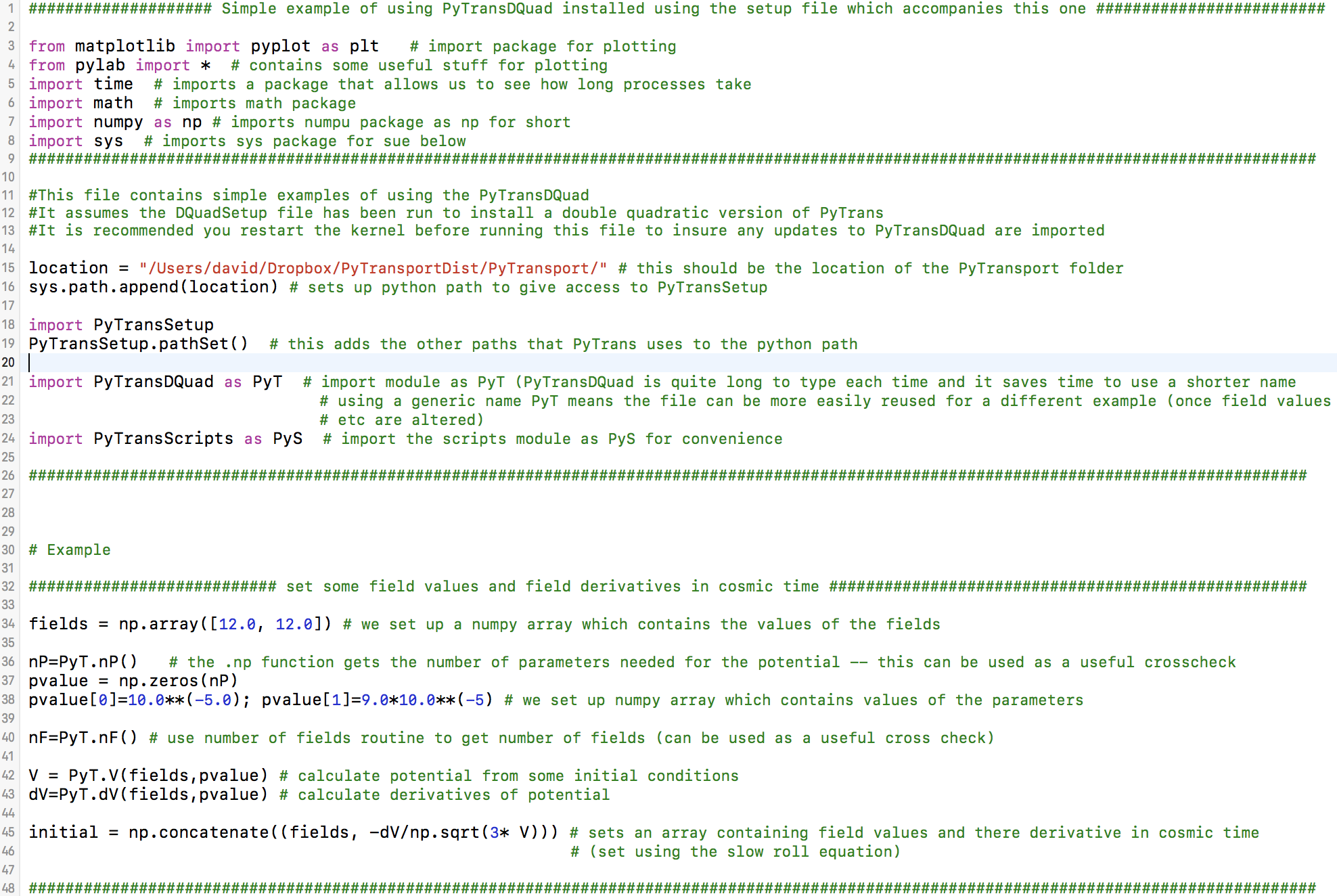}
\end{figure}

\noindent This screen shot is of the start of the file {\it Examples/DoubleQuad/SimpleExample.py}. Of course we will usually want 
to use the module for more sophisticated tasks. Appendix~3 contains a summary 
of all the functions  available within the { PyTrans***} module. We will see the use of a number of the more sophisticated functions in the Examples section.

\section{Examples}
\label{Examples}

\subsection{Double quadratic}

\begin{figure}[H]
\centering
\includegraphics[width=7cm]{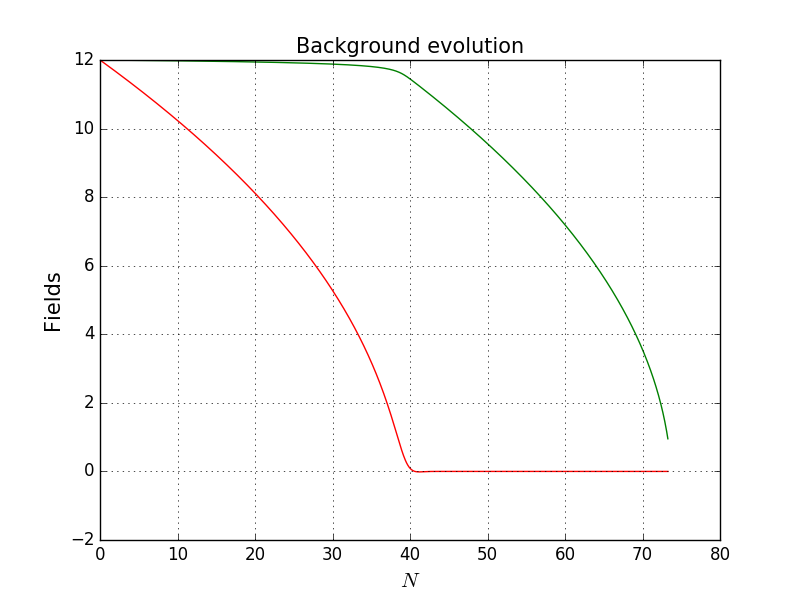}
\caption{\label{double1}
Background evolution for double quadratic potential}\label{shot3}
\end{figure}

First lets continue with the double quadratic example using more of the functions 
available from the compiled module. In the screen shot below we use the background evolution 
function to calculate a fidicual background trajectory in field space using the array we set up in the last 
part of the example as initial conditions. Here and for all the functions and output of 
the PyTransport package e-folds ($N$) are used as the time variable.
The function used to calculate the background evolution is the {\it PyTrans.backEvolve} function. 
Details of the format it outputs the background evolution in can be found in Appendix~3. Essentially it provides 
information about the fields and their rate of change (in cosmic time) at every e-fold value given by 
the array $t$. 
In addition to this array and the initial conditions, 
one must provide the parameter value used to define the potential, and the absolute and 
relative tolerances to be used by the integrator (this argument has been added 
in the second version of the code). 
The final argument of this function indicates whether the evolution should terminate at the end of inflation 
(and has also been added in the second version of this code).
If set to true data is only returned up to the first value of $t$ that is past the end of inflation if that point is reached (note it does not 
find the exact end of inflation, and the last value returned will only be close to the end of inflation if the array $t$ 
finely samples the evolution). If set to false, the code will attempt to generate output for all the times contained in $t$. If the 
last entry is long past the end of inflation the code could take a long time to run (or crash) as typically 
the number of oscillations in field space grows exponentially with $N$.
The output is plotted shown in Fig.~\ref{double1}. 
\begin{figure}[H]
\centering
\includegraphics[width=18.cm]{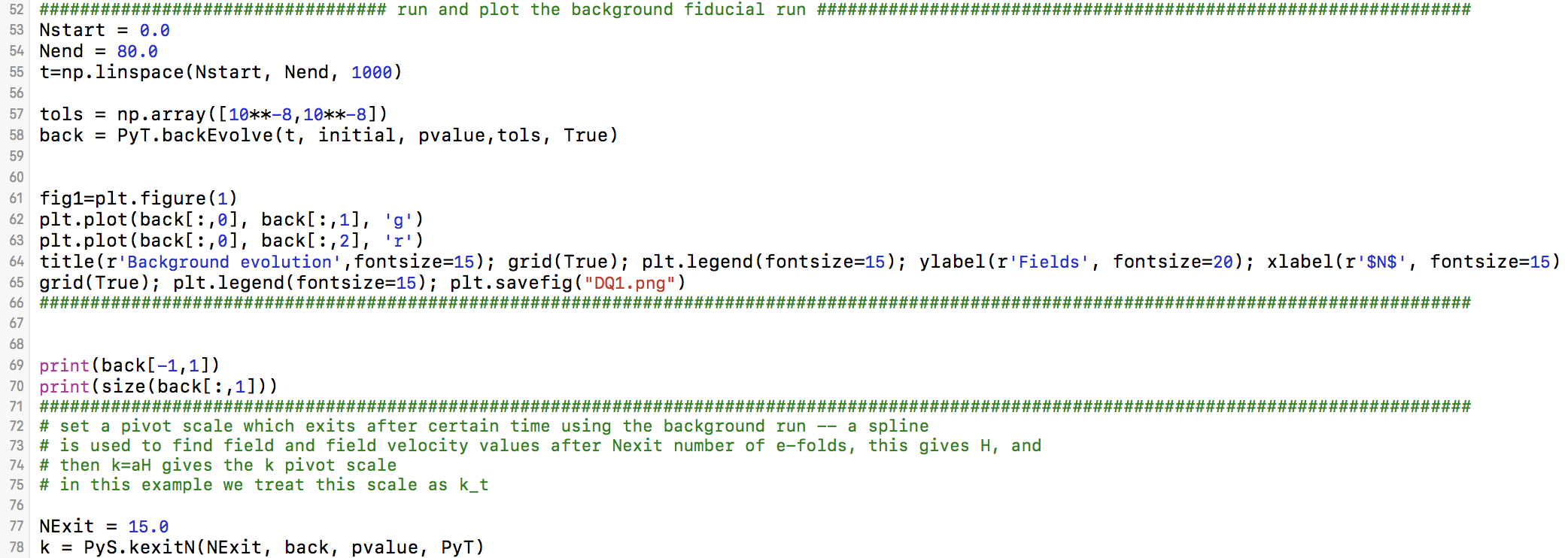}
\end{figure}

\begin{figure}[H]
\centering
\includegraphics[width=7cm]{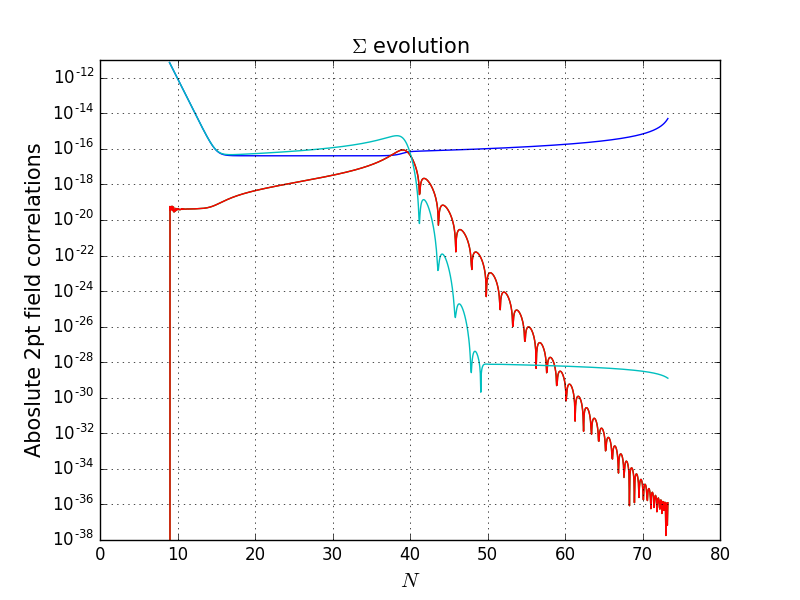}\includegraphics[width=7cm]{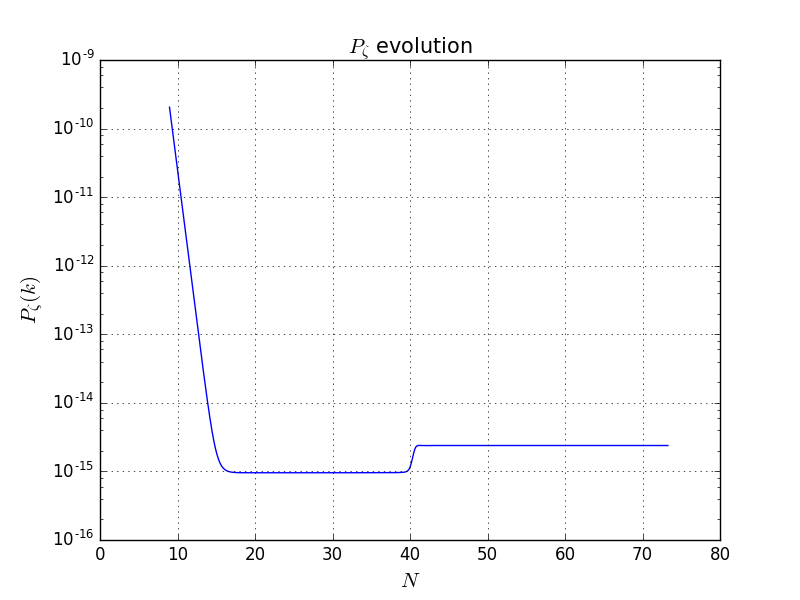}\
\includegraphics[width=7cm]{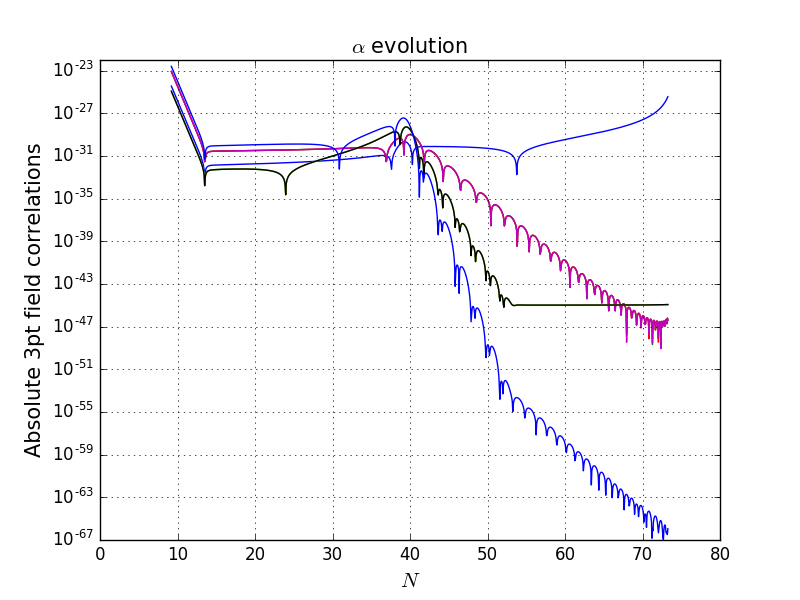}\includegraphics[width=7cm]{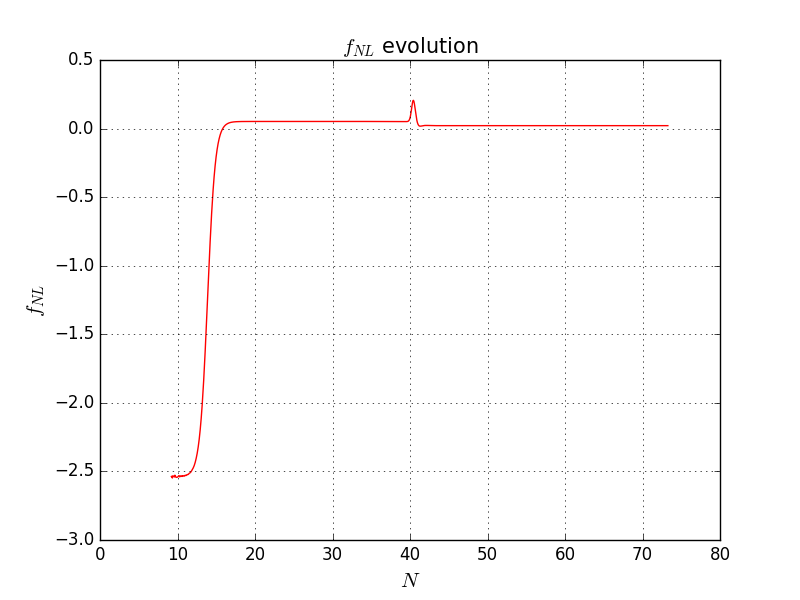}\
\caption{\label{double2}
Evolution of correlations for double quadratic potential from $6$ e-folds inside the 
horizon for a Fourier mode
which crosses the horizon at $N=15$}\label{shot3}
\end{figure}

Then we run the two point evolution function to calculate the evolution 
of $\Sigma$ and the power spectrum of $\zeta$ for a $k$ mode which crossed the horizon 
15 e-folds into this fiducial run using the {\it PyTrans.sigEvolve} function. 
We plot the correlations and cross correlations of the fields in Fig.~\ref{double2}. 
We repeat for a neighbouring $k$ to give us a crude estimate of the spectral index,  $n_s$. 
Finally, we use the {\it PyTrans.alphaEvolve} function to calculate the evolution of the field space three-point function 
  and the bispectrum of $\zeta$  for a set of three $k$s.
We plot the three-point correlations and cross correlations of the fields in Fig.~\ref{double2} and also 
the evolution of the reduced 
bispectrum, $f_{\rm NL}$. In the plots it can clearly be 
seen that the heavier field drops out of the dynamics at around 40 e-folds. At this 
point the system becomes adiabatic and $\zeta$ and its statistics become constant.
A screen shot of the code which does all this from the {\it SimpleExample.py} file is below:
\begin{figure}[H]
\centering
\includegraphics[width=18cm]{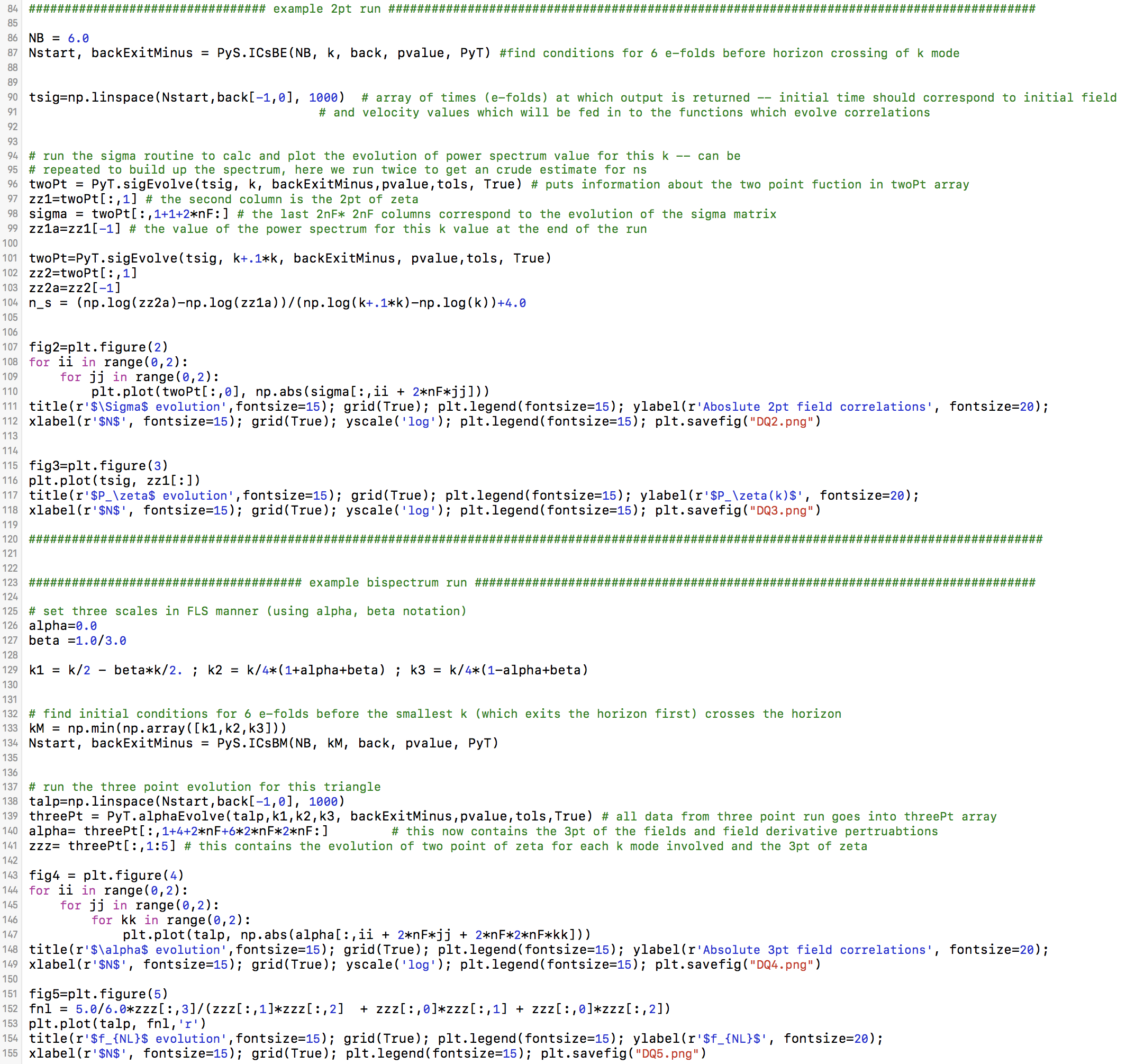}
\end{figure}

\noindent There are few things to note from this script. 

First we note that  
it is important that we fix our final time sensibly. 
If we were to fix it after the end of inflation, when the lighter field 
oscillates indefinitely about its minimum, then the code would become very slow, especially 
when it evaluates the 
correlations (since the the field correlations also oscillate a lot in 
response). 
This can be achieved by using a time array which 
runs up to the last output time of the background evolution (if, as in this example, we asked the background evolution 
to terminate at the end 
of inflation).  This final time may not be what we require (if for example the moments of $\zeta$ becomes constant 
before the end of inflation), and in this case we could use an earlier time. 

Next we note the use of a function within {\it PyTransScripts.py} which 
finds the k value which corresponds to an exit time of 15 e-folds after the start of the fiducial run, the {\it PyTransScripts.kexitN} 
function. This function 
uses the background trajectory and Python spline routines to find this k.
We also note that it is essential that we run the two and three point correlation evolution 
from a time the k mode of interest is deep inside the horizon. In the script, we calculate this time 
as well as the field and field velocity values at this time 
(which are then fed into the two and three point evolution routines) using another function from the 
scripts module, the {\it PyTransScripts.ICsBE} function. 
To use this function we need to specify the number of e-folds we require before horizon crossing, and here we specify $6.0$. The function returns a time roughly 6 e-folds before the horizon crossing time and a numpy array containing the fields' values and velocities at that time. 
It is important to point out that this function is only approximate in the sense that it 
simply finds the first value of $N$ in the background array (back) 
which is before the specified number of e-folds before horizon 
crossing, and returns this value and the corresponding field and field velocities at this time. 
It therefore requires 
the background fiducial evolution which is fed into it to be finely sampled 
(10 points per e-fold is a rough guide) to be accurate.

Finally, we note 
that within {\it PyTransScripts.py} we also include the related functions  {\it kexitPhi} which finds the k value which 
crosses the horizon at a particular field value, and and {\it ICsBM} which finds initial conditions a 
fixed time before the ``massless condition" which is discussed further in section~\ref{heavy}.  These functions 
and all others in this module are detailed in appendix 4.


\begin{figure}
\centering
\includegraphics[width=8cm]{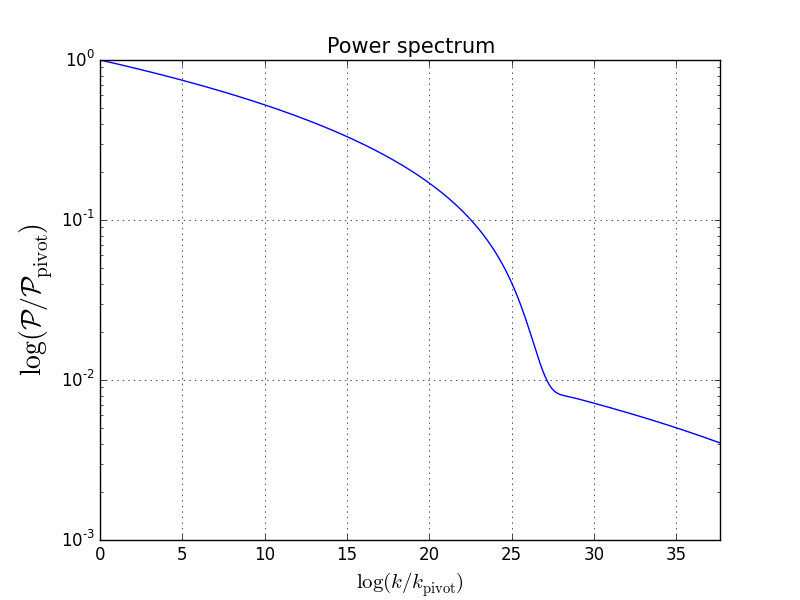}\includegraphics[width=8cm]{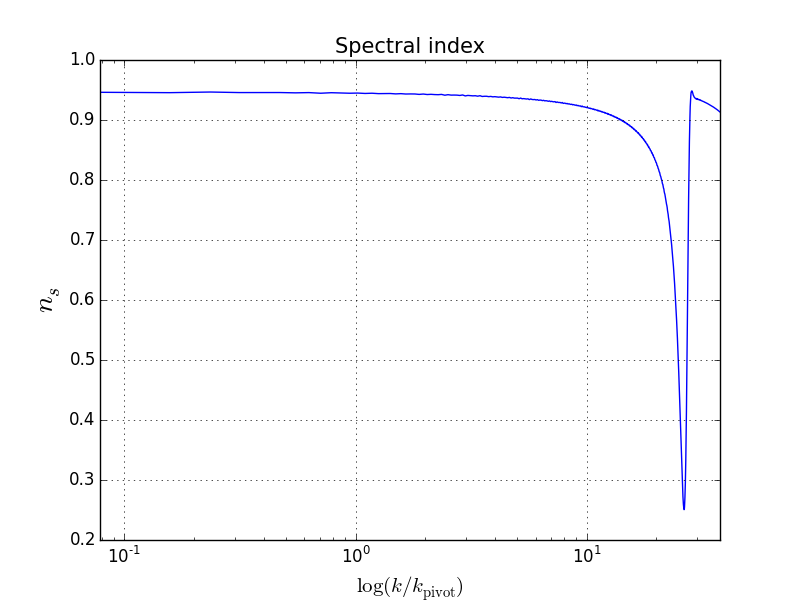}
\caption{The power spectrum and $n_s$ in the double quadratic model. \label{power}}
\end{figure}

Calculating the value of $n_s$ in the manner presented 
here is clearly a bad way of doing things, since it involves using only two points 
in the power spectrum to calculate a derivative, and the step between them is arbitrarily chosen. 
An alternative is to calculate the power spectrum 
around a given exit time using a number of points, and to fit a spline to it and differentiate. We now generate the 
power spectrum for this model with the 
following script which fits a spline to the entire spectrum, differentiates and produces $n_s$ at every value of 
$k$ over roughly 30 e-folds, the results are plotted in Fig.~\ref{power} (the start of the file is identical to that above):
\begin{figure}[H]
\centering
\includegraphics[width=18cm]{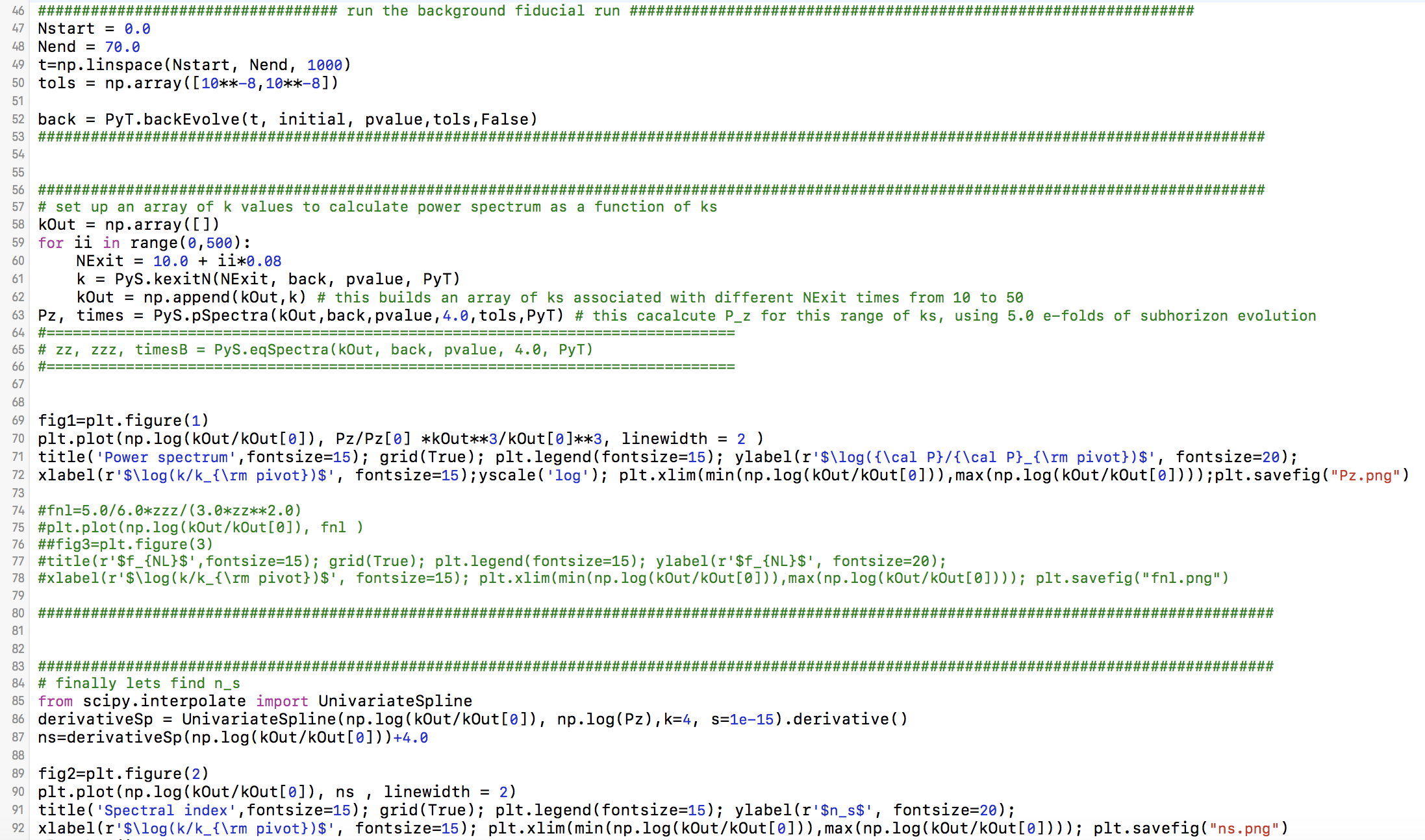}
\end{figure}

\noindent In this script we used the {\it PyTranScripts.pSpectra} function to generate the power spectrum over a 
range of $k$s. This function essentially just loops over calls to the compiled function which evolves the two-point function.

Next we wish to calculate the bispectrum. Here we first we calculate the bispectrum 
in the equilateral triangle configuration as a function of the $k$ values we calculated the power spectrum 
for above. Then 
we generate (and plot using a separate plots file) a slice through the bispectrum for a given 
$k_t$ as a function of the $\alpha$, $\beta$ variables defined such that $k_1 = k_t/2 - \beta  k_2/2$, 
$ k_2 = k_t/4*(1+\alpha+\beta)$ and $k3 = k_t/4*(1-\alpha+\beta)$. 
We use two separate scripts for each of these tasks (and the plots file) which are pasted below, 
both use MPI to speed 
things up. They should be called 
using the command ``/usr/local/bin/mpiexec', '-n', '10', 'python', 'MpiEqBi.py' '' 
 (for the equilateral run, where the first part 
should be replaced 
with your location of mpiexec if different, and 10 replaced by the number of processes one desires to call): 
\begin{figure}[H]
\centering
\includegraphics[width=18cm]{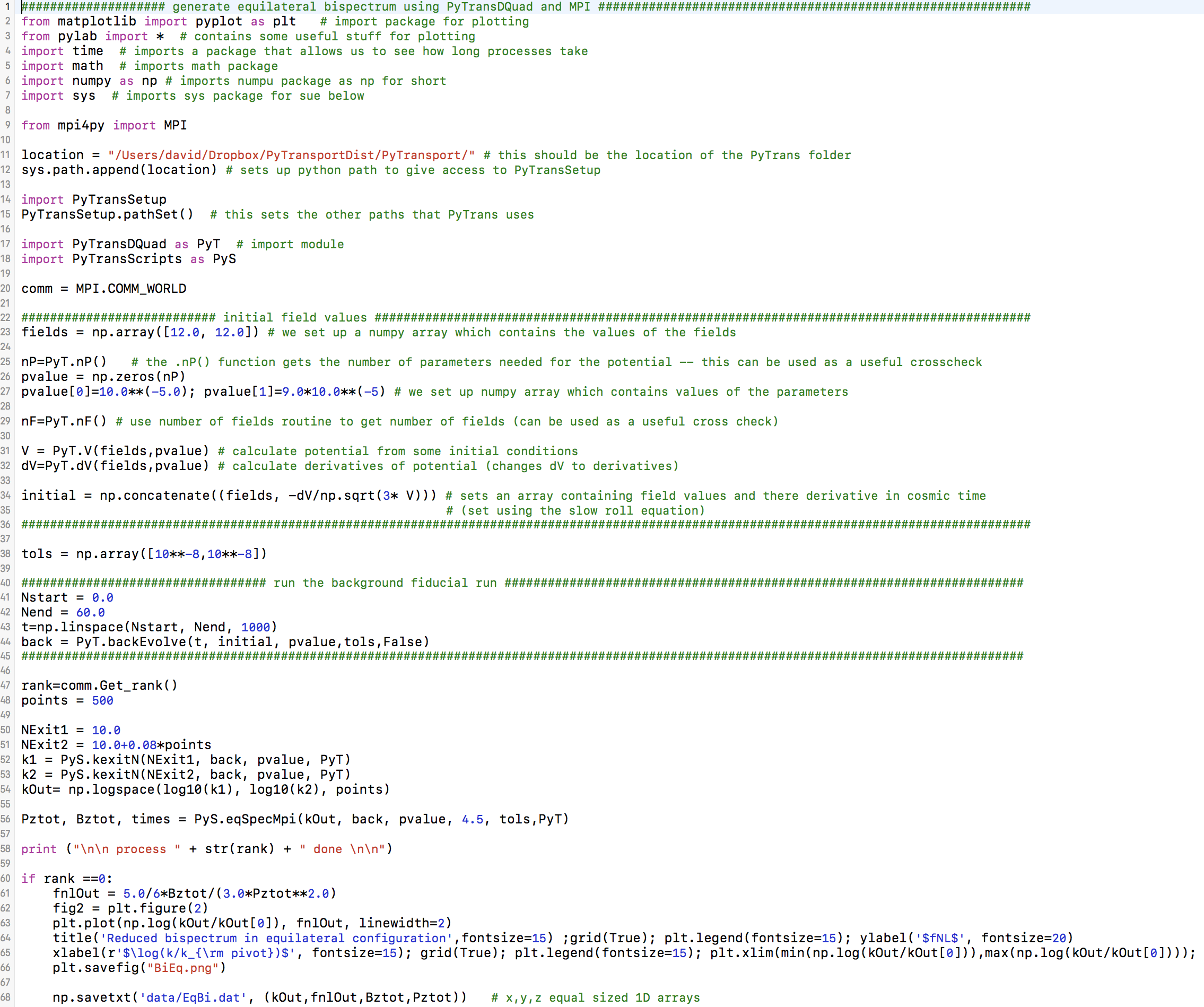}
\end{figure}

\begin{figure}[H]
\centering
\includegraphics[width=18cm]{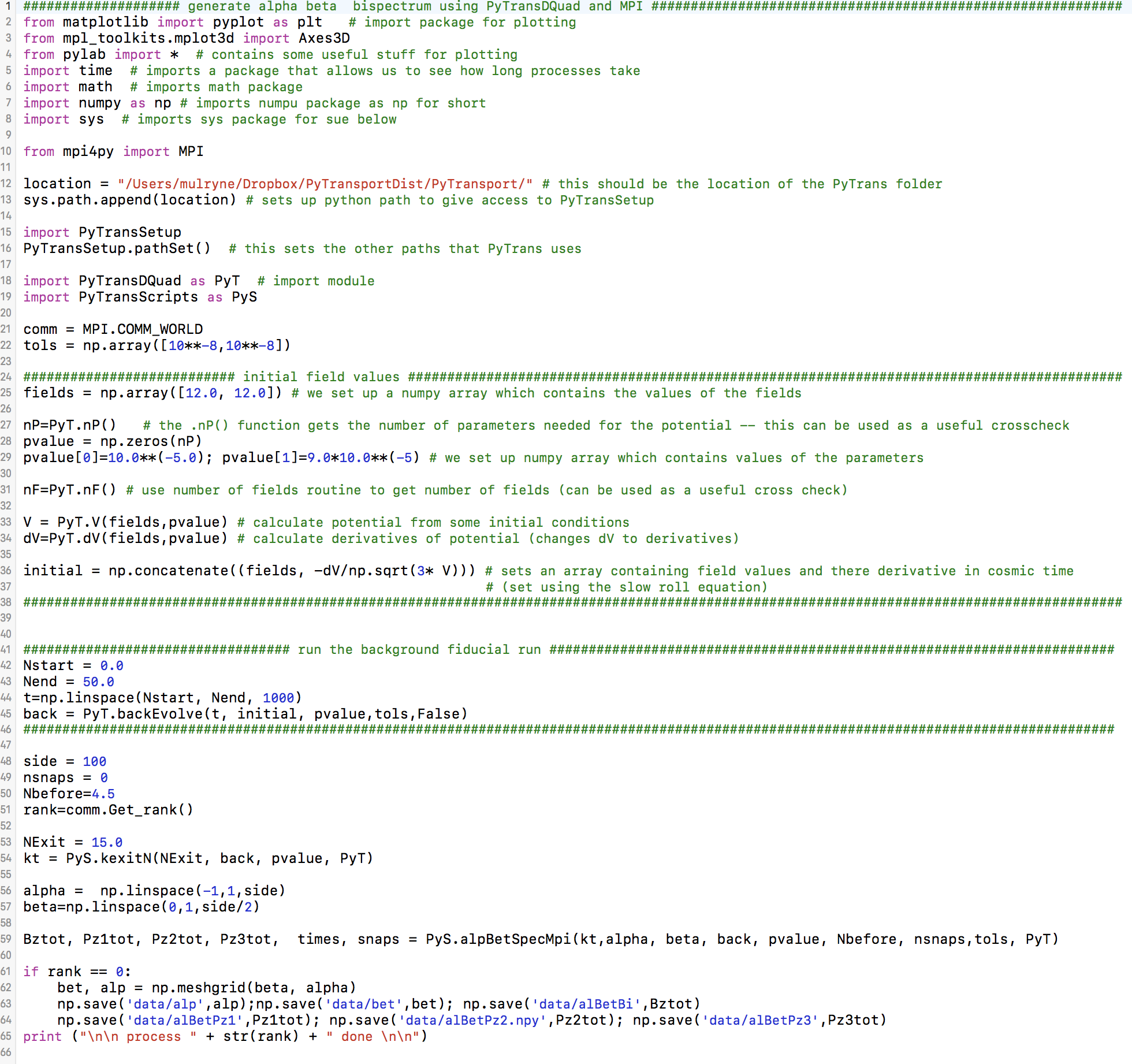}
\end{figure}

\begin{figure}
\centering
\includegraphics[width=8cm]{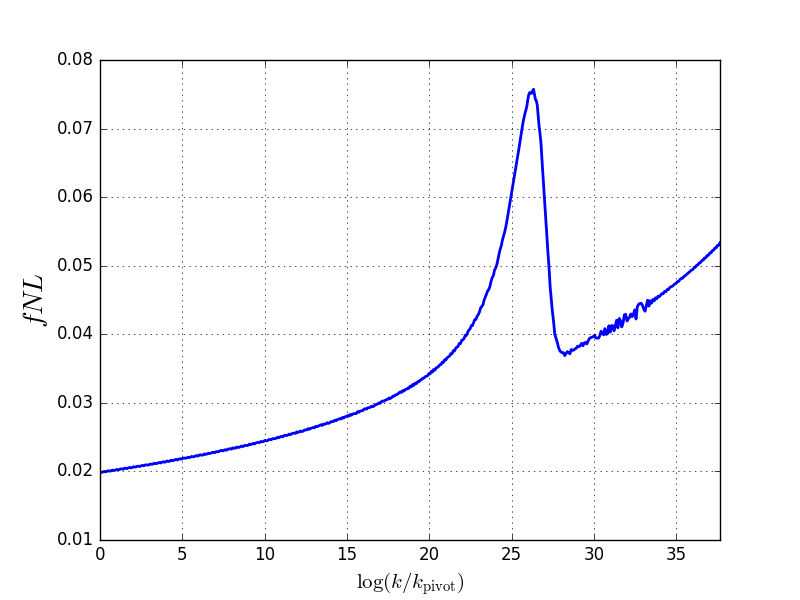}
\caption{The reduced bispectrum in equilateral configurations for the double quadratic potential.\label{eqi}}
\end{figure}

\begin{figure}
\centering
\includegraphics[width=9cm]{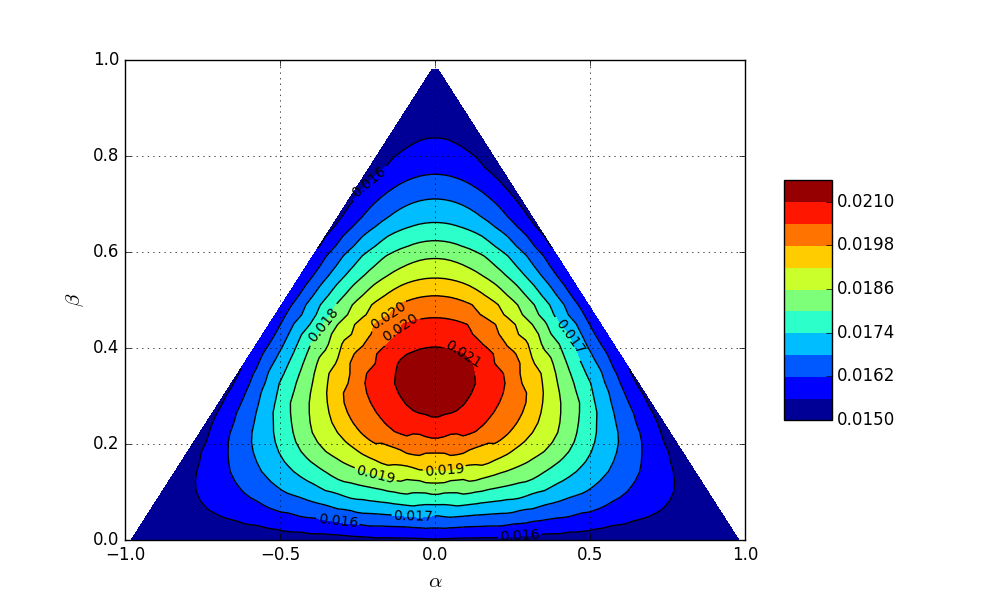} \hspace{-1cm}
\includegraphics[width=7.5cm]{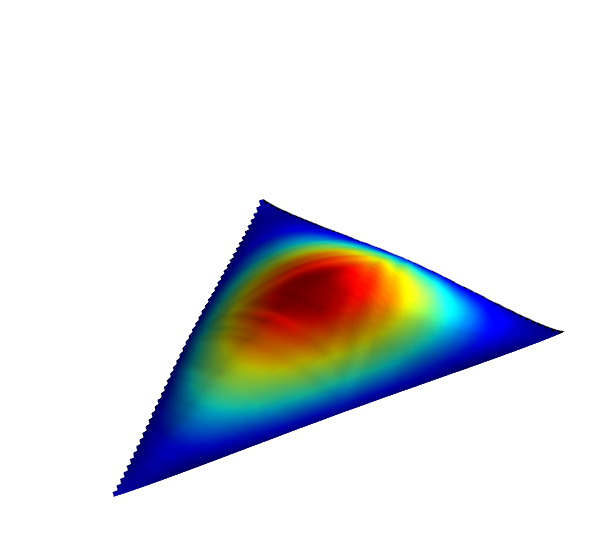}
\caption{A slice through the reduced bispectrum for a mode exiting after $15$ e-folds for the double quadratic potential run discussed in the text. \label{alpbet}}
\end{figure}

\noindent The results are in Figs.~\ref{eqi} and \ref{alpbet} respectively. 
Note that in all these scripts we use more functions available from the { PyTransScripts} module whose function 
should be self evident and which are described in full in appendix 4.
If one didn't wish to use MPI the only change needed would be to call the function {\it alpBetSpec} rather than 
{\it alpBetSpecMpi} from the scripts module, and to remove MPI related lines and reference to 
rank etc.
When using MPI we recommend calling more processes 
that the system has cores. This is because we have not implemented sophisticated load sharing, and since 
some ranges will be faster to evaluate than others, if 
the number of processes is larger than cores, the cores that have capacity first will end up running more processes, 
sharing the load in a simple way. In this example we also save the data at the end for future use. This can be done 
simply for any numpy array in various ways and then read back into Python easily. For reasons of simplicity and flexibility 
we leave data management up to the user, but note that Python is a powerful tool for this purpose.

If we want to generate the full bispectrum we would simply loop over the {\it alpBetSpec} or {\it alpBetSpecMpi}
function for many $k_t$.

\subsection{Double quadratic with a field-space metric}
\label{NCexample}
To demonstrate the evaluation of a non-canonical inflation model with a curved field-space metric 
we extend the previous example to include a non-trivial metric. 
We retain the potential from Eq.~(\ref{dubquadpot}), but now consider the model to have the field metric
\be
\label{GNC}
G_{IJ}=\left(\begin{matrix}
R_0^2 & 0\\
0 & R_0^2\sin\phi^2
\end{matrix}\right)\,.
\ee
Previously we needed to create a Python module for a model with the double quadratic potential (that was canonical),
now a Python module must be created for a model with this potential and the field metric. This is achieved by writing out the setup script we had before but including a $G$ matrix which encodes Eq.~(\ref{GNC}) in SymPy notation. 
This is then included as a final argument in the {\it potential(V,nF,nP,Simple,G)} function which sets up the files needed to compile this model 
into a python module (this final argument is optional, if not included then model must be canonical). The fourth 
argument is also optional and we didn't discuss it when dealing with the previous canonical example. It tells the routines 
which use sympy whether or not to attempt to simplify the expressions using SymPy's simplification routines. By default it is set 
to False for reasons discussed in Section\,\ref{pcf} (this ability to switch simplification off has also been added to the 
new version of the code). 
Finally, the function {\it compileName(`DQuadNC','True')}, or {\it compileName3(`DQuadNC','True')} if the user is using Python 3, is 
used to create a python module {\it PyTransDQuadNC}, which can be used to study this model. The final argument in the {\it compileName} function is optional, and care must be taken to include it and set it to True if the model has a 
non-trivial field space metric. 
If set to true the then the code that deals with the compilation is pointed to different \CC files 
which contain the more complex equations needed deal with models of inflation
with a non-trivial field space. Below is a 
screenshot of setup file for this model. This example can be found in the {\it Examples/DoubleQuadraticNC/} folder with 
some simple scripts similar to those above for the canonical case. The complied model can be used in exactly the 
same way as for canonical example.
\begin{figure}[H]
\centering
\includegraphics[width=18.0cm]{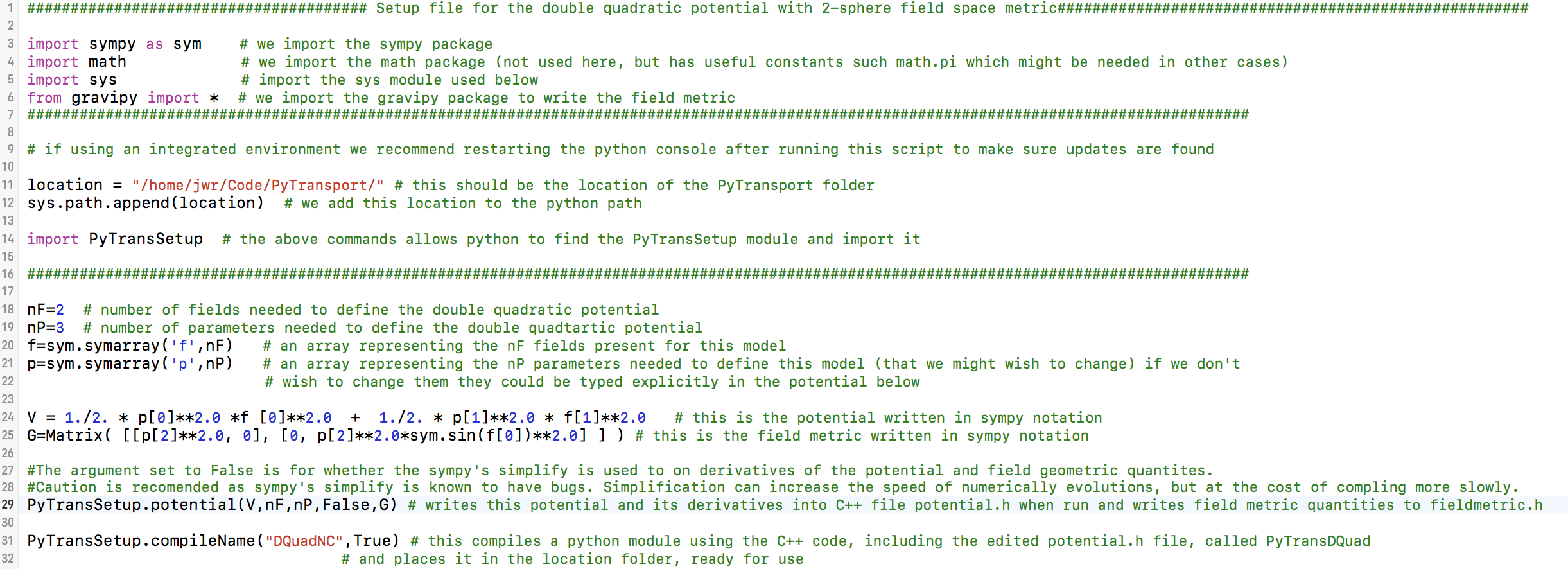}
\end{figure}
\subsection{Heavy field examples}
\label{heavy}

In the previous examples both fields which played a role in the dynamics 
were light (at least at the start of the evolution). Interesting dynamics 
can also occur when the field orthogonal to the direction of travel 
in field space is heavy, if 
the field trajectory curves. In this kind of example it is imperative that the initial conditions for the evolution of the two and three point functions 
are set when the $k^2/a^2$ term in the equation of motion for the scalar field perturbations 
dominates over the mass squared of the heavy 
field. This is a requirement for our initial conditions to be accurate as discussed in Ref~\cite{Dias:2016rjq}. 
There is a script in {\it PyTransScripts.py} which 
will achieve this, the {\it ICsBM} function. This finds initial conditions a user specified number of e-folds before the 
massless condition where $k^2/a^2=M^2$ (where M is the largest eigenvalue of the mass matrix). 
There is also the function {\it ICs} which evaluates initial conditions using {\it ICsBE} (the before horizon exit 
function) and {\it ICsBM}, and takes the earliest one. The power spectrum and bipsectrum routines use this latter function.

One example is the potential:
\be
V = \frac{1}{2} m^2_\phi \phi^2 + \frac{1}{2}M^2 \cos^2\left ( \frac{\Delta \theta}{2} \right) \left [\chi - (\phi - \phi_0 ) \tan \Xi \right]^2 
\ee
where 
\be
\Xi = \frac{\Delta \Theta}{\phi} \arctan[s (\phi - \phi_0)]
\ee
which is from Ref.~\cite{Gao:2013ota} and is in the  {\it Examples/LH/} folder with some 
simple scripts to those discussed above for the double quadratic potential for users to play with. 
This example is also discussed at length in Ref.~\cite{Dias:2016rjq}.

\subsection{Further examples}
\label{other}

Also in the {\it Examples/} folder is another light field example with more interesting dynamics than the double quadratic 
example which we refer to as the axion quartic model. This example is again accompanied 
with scripts and plots for users to explore and was also discussed in Ref.~\cite{Dias:2016rjq}. It is in the {\it QuadAx} folder, and has the potential
\be
V = \frac{1}{4} \lambda \phi^4 + \Lambda^4 \left ( 1 - \cos \left ( \frac{2\pi \chi}{f} \right) \right )\,.
\ee

Finally in the {\it Examples/} folder and discussed in  Ref.~\cite{Dias:2016rjq} is a single field example with a step in the potential:
\be
V = \frac{1}{2} m^2 \phi^2 \left (1+c \tanh\left(\frac{\phi-\phi_0}{d} \right)\right)
\ee
which is in the folder {\it SingleField/} and was discussed in Refs.~\cite{Chen:2006xjb,Chen:2008wn}, as well as in Ref.~\cite{Dias:2016rjq} .

\section{Things that can go wrong}

While using the PyTransport package some issues have presented themselves which it might be useful 
for new users to know about. Many more have been ironed out, but it is neither practical nor desirable to make the code 
fully immune to misuse. Below we detail a few common problems we have faced.

\subsection{Potential computation fails and problems with simplification of expressions}
\label{pcf}
This is the most severe bug/problem we have found, but also the least common. 

The function {\it PyTransScripts.potential()} takes a potential V written in SymPy format (and a field 
space metric, if the model is not canonical) and 
calculates and then simplifies related functions such as the derivatives. To do so it uses the function {\it sympy.simplify}. 
As discussed at 
this \href{http://docs.sympy.org/latest/tutorial/simplification.html}{reference}\footnote{http://docs.sympy.org/latest/tutorial/simplification.html}, however, this is not always 
the best way to simplify an expression and can take some time to complete the simplification 
for complicated functions. For example the simplification process takes a relatively long time (tens of seconds) for the heavy 
field model 
above.

We found a more serious problem occurred when looking at the example:
\be
V = V_0 \left (10 - \sqrt {2 \epsilon_s} \arctan \left ( \frac{\phi}{\chi} \right ) +    \frac{1}{2} m_\chi^2 
\left ( \sqrt {\left (\chi^2 + \phi^2 \right )} - 2 \right )^2 \right)
\ee
which represents a semi-circular valley in field space.
For this example, there appears to be a 
{\it simpy.simplify} bug which made it crash with 
the potential written in the form given above, when the powers where written as doubles -- i.e. as 2.0 -- rather than simply as 
2. The problem appears when taking cross derivatives. 
Uncommon errors such as this are something users might need to watch out for. 

We have also noticed that it is helpful to write powers as integers (such as 2, rather than 2.0) in general, as this can result in 
much shorter expressions after simplification. If analysing a complex model, or if an error is encountered, the user can 
inspect the potential.h file or the fieldmetric.h file to see what expressions Python has generated. The location of these files is 
discussed in appendix~1.

Because of these issues simplification is by default switched off, but can be useful to speed up the code.

\subsection{Make sure latest module version is imported}

A more common problem is that if we wish to update a complied PyTrans*** module, for 
example after altering the tolerances, then after recompiling the module we need to 
ensure the new module is imported. The only reliable way to do this seems to be to restart or open 
a new Python session
 and then use the import command. If working in the Canopy editor, for example, this can be achieved by selecting 
 ``restart kernel" from the ``run" menu. 

\subsection{Selecting the absolute and relative tolerances}

The evolution of the three point function  can be a numerically intensive task, requiring high numerical 
accuracy. The question arises how low (the lower the higher the accuracy) do we need to set numerical 
tolerances. This question can't be answered absolutely, and must be dealt with on a model by model basis. Models 
with finer features in the potential, or in which the excitation of the two and three point function 
occurs on sub-horizon scales will require lower tolerances (high accuracy). Models which produce a small signature may also need 
higher accuracy to resolve the true answer from noise than models which produce a large bispectrum. 

Convergence is the key criterion in selecting tolerances. If one calculates the evolution of the three-point 
function of $\zeta$ for a example run with a given potential, reduce the tolerances and run again without the answer 
changing in any significant manner, then the tolerances are likely to be sufficient.  As a rule of thumb 
$10^{-8}$ for the absolute tolerance and $10^{-8}$ for the relative tolerance is usually sufficient. 
However, some examples do require lower values, while for reasonably accurate results for simple models 
one can sometimes get away with higher values. As the values 
are lowered,  the code takes longer to run and eventually will fail (described below). Therefore, 
there is significant benefit for picking a required accuracy which is sufficient for the task, but not one which is too 
stringent. Those with experience of solving ODEs numerically will be familiar with this problem, which is of course 
more general than the specific integration at hand. One can determine the accuracy needed for a given model 
by examining the output for various choices for a small number of runs and checking for convergence.

\subsection{Integration stalling or failing} 

The consequences of picking a tolerance that is too demanding can be that the integrator will not finish. 
But this can also be a consequence of setting silly initial conditions or parameter choices. It is always a good idea 
once you begin with a new model of inflation to build up gradually. First integrate the background. Even here it 
is possible for the code to take a long time. For example, if the final time is after the end of inflation the code will 
try and  track all the oscillations of the field. As these increase exponentially with e-fold it 
can be very time consuming. 
Once the background is giving sensible output, move onto integrating the correlations, 
initially just for a 
single $k$ value of the power spectrum or single triangle of the bispectrum. If 
everything looks good then run over many values to calculate the power or bi-spectrum. If you feel you are waiting too long try just asking 
the code to evolve 
a short e-folding time (.1 say). Typically a single triangle of the bispectrum evaluated before the end of 
inflation will take 
from between a fraction of a second to half a minute to run with $4-6$ efolds of 
sub-horizon evolution, depending on the required accuracy. Heavy field models however typically have many more sub-horizon efolds given that the massless condition is already met only deep inside the horizon.  If the run time seems to be much longer than you expect, double check 
you are working  
with the correct potential for your initial conditions and parameter choices.

\subsection{Integration failing}

If the code can't reach the accuracy demanded by the user the rkf45 routine will stop running and issue an error 
message that the required accuracy couldn't be reached. 

\subsection{Not enough efolds before horizon exit}

A requirement that needs to be met in order to get accurate power spectra and bispectrum is that all the k values 
involved in a given correlation must be sufficiently deep inside the horizon initially for the 
initial conditions to be accurate. Since the functions which find the initial conditions for the two and three point evolutions take a background 
trajectory as their input, this trajectory must have enough e-folds prior to the exit of the k values of 
interest so that the correct initial conditions can be found. 

Moreover, we must choose how many e-folds these k values stay sub-horizon for. As described above we can either 
measure backwards from horizon crossing itself or from the massless condition (only suitable for models with heavy fields) or pick 
the earliest of the two conditions. 
Normally 4-5 e-fold from these points is about right. But as with the 
setting of tolerances, the only way to tell for sure is to demand convergence. Too little run in 
time will lead to spurious oscillations in the spectra.

\subsection{Not enough efolds after horizon exit}

For single field models or effective single field models, such as models additional heavy fields, the correlations of 
$\zeta$ become conserved after horizon crossing. Likewise if there are multiple additional light 
fields which then decay, leading to 
adiabatic evolution, $\zeta$ also becomes conserved. In the former case conservation only 
occurs a few e-folds after 
horizon crossing, and in the latter the process also takes a few e-folds to complete. 
We must ensure, therefore, that the statistics are 
evaluated at a time when $\zeta$ has become conserved if that is what is intended, and so our evaluation time 
must be sufficiently far after horizon crossing or field decay.

\subsection{Computer crashes and data loss}

The { PyTrans***} compiled module returns data in memory to Python. Moreover, the functions provided 
to calculate power spectra and bispectra which loop over the complied module also return 
arrays of data in memory to the calling process. This is true even for the MPI functions. If anything goes wrong before these functions complete the data is lost. For example if the computer crashes, or if using distributed computing any 
one of process crashes. These modules can be easily edited by users to instead write data to disk 
periodically, particularly advisable for distributed computing. We have not 
included this functionality to avoid overly complicated functions, and in any case, we anticipate that 
different users will have different needs, but it is an issue to look out for.

\section{Summary}

We have presented the PyTransport package for the calculations of inflationary spectra. This package 
compliments and extends currently available tools, and is also complimentary to two related packages 
mTransport \cite{Dias:2015rca} and CppTransport \cite{Dias:2016rjq,Seery:2016lko}  all described at this \href{https://transportmethod.com}{website}\footnote{https://transportmethod.com}. In its most recent version it has also now been updated to include 
functionality to deal with models that have a curved field space metric \cite{xxx2}.

Through use of a detailed example we have shown how PyTransport can be used in practice. We have also summarised the structure of the code, with some more details provided in the appendices.

\begin{acknowledgments}
DJM is supported by a Royal Society University research fellowship. JWR acknowledges the support of a studentship 
jointly funded by Queen Mary University 
of London and by the Frederick Perren Fund of the University of London. This work would not have been possible without 
the related collaboration with David Seery, Mafalda Dias and Jonathan Frazer \cite{Dias:2016rjq}. We also 
thank Sean Butchers and David Seery for helpful cross checks  made between PyTransport and CppTransport.
\end{acknowledgments}

\section*{Appendix 1:  Base code description}
We now give a brief description of the \CC \S  code which is the core of PyTransport. 

\noindent The folder  {\it PyTransportDist/PyTransport/CppTrans/} contains the base \CC \S code which the complied Python module uses.
Inside lies another folder {\it PyTransportDist/PyTransport/CppTrans/NC/} which contains the 
five header files from the main directory but modified for non-canonical models. If a model contains a field metric the {\it PyTransSetup.compile()} function will redirect the path to the files within this directory (the final argument needs to be set to True
for this to happen). 
The reason for this separation is to optimise the speed of computations when dealing with only canonical models.
The biggest file is {\it model.h}. This contains the model class, which 
defines the properties of an inflationary model. It contains member functions which allow the calculation of important 
properties of the inflationary model, from simple objects such as the value of the Hubble rate, to more complex ones such 
as the 
$A$, $B$, $C$ tensors which define the third order action, as discussed in Refs.~\cite{Dias:2016rjq,xxx2}. It also calculates the ``N" tensors, 
mentioned in the main text,
that determine how field fluctuations are related to $\zeta$. 
The equations of motion for the transport system are written in terms of ``$u$" tensors (see section~5 of Ref.~\cite{Dias:2016rjq} and Ref.~\cite{xxx2} for the curved field space generalisation), and there is also a member function of the model class which calculates these tensors. As noted in the main text 
we rescale the field velocity perturbation for performance 
reasons using the {\it scale} member function of the model class. This rescaling appears in the 
member functions which calculate $u_2$ and $u_3$ tensors and the $N$ tensors, altering them slightly 
from the from given in Ref.~\cite{Dias:2016rjq}.
The output of these functions is also divided by the Hubble rate, changing the 
time variable from cosmic time (which the equations are written in in Ref.~\cite{Dias:2016rjq}) 
to $N$, e-fold time.  The scaling we choose essentially means the initial conditions for 
correlations of the velocity perturbations
are of the same order of magnitude as correlations of field perturbations, helping the performance of the integrator. 

The {\it moments.h} file contains classes which define properties of the two and three point functions of field perturbations. 
This class is not extensively used by the Python code, but the constructor of these classes allows an instance of the 
two/three-point function to be created containing the values these objects take deep inside the horizon -- ie the initial 
conditions needed by the code. 
Once again these initial conditions are rescaled using the {\it scale} member function of the model class so that the initial 
conditions match the evolved 
correlations.

The {\it potential.h} file contains a class which defines the properties of the potential, and importantly has member functions which provide the derivatives of V. There is no barrier to changing this file by hand for a new potential. But the directory also contains {\it potentialProto.h}, which is a file that contains the skeleton of the {\it potential.h} file, and is used to automatically generate a potential.h file by the {\it PyTransSetup.potential()} function, detailed in appendix 1. 

The {\it fieldmetric.h} file contains a class with functions containing the tensors defining the field-space such as the field-metric, Christoffel symbol, Riemann tensor and it's covariant derivative. Like the {\it potential.h} file, the content may be changed by hand. Likewise a skeleton file {\it fieldmetricProto.h}, used to generate the {\it fieldmetric.h} file when a flag is called in the {\it PyTransSetup.potential()} function, is within the same directory.

The {\it evolve.h} file contains functions which define the evolution equations for three systems in the form needed 
by the rk45 ODE stepper we use: the background only system, the two point function  
coupled to background, and three point coupled to two point coupled to background. The system of equations of the latter two systems 
are given 
in Eqs.~(5.5) and (5.16) of Ref.~\cite{Dias:2016rjq}). These equations use the ``u" tensors which are also described in 
section~5 of Ref.~\cite{Dias:2016rjq} (and Ref.~\cite{xxx2} for the curved field space generalisation) and 
which are calculated by the {\it model.h} class.  

Finally there is a folder called rk45 stepper
which contains the code for the ODE solver routine we use.

\section*{Appendix 2: Setup functions}

The functions which are part of the { PyTransSetup} modules.

\begin{itemize}
\item    {\it \bf PyTransSetup.directory(NC)} takes one argument ``NC", a boolean value indicating whether the model is non-canonical {\color{blue}\bf{True}} or not {\color{blue}\bf{False}} (by default set to false).
The function automatically edits the \CC \S file {\it PyTrans/PyTrans.cpp} so it knows about the absolute location in the file system of the \CC \S files  {\it potential.h}, {\it model.h} and {\it moments.h}. If NC={\color{blue}\bf{True}} the locations of files within the {\it CppTrans/NC/} are instead used. A user shouldn't need to call this function themselves, it is called by the compile functions below.

\item    {\it \bf PyTransSetup.pathSet()} takes no arguments and adds the location of {\it PyTrans} compiled module to the Python path. This function was called in the examples above.


\item    {\it \bf PyTransSetup.potential(V,nF,nP,simple,G)} takes a expression in SymPy format ``V", which is a function of two SymPy arrays p and f, integers ``nF" and ``nP " which are the length of the arrays p and f, ``simple" an optional Boolean value (by default set to false) which determines whether SymPy's, and an optional SymPy matrix ``G", which is also a function of p and f (by default set to ``$0$", a flag that prevents the fieldmetric from being calculated).
simplify function is used to simplify the expressions stored\footnote{User discretion is advised when setting simple = {\color{blue}\bf{True}}. While often it can speed up computation time when running evolution functions, since simplify attempts multiple types of simplifications until it determines which is best, it may unnecessarily slow down compilation time. In addition some other pitfalls of the simplify function are discussed in \ref{pcf}}. The function then writes this potential and its derivates into the {\it PyTransport/CppTrans/potential.h} file (using the {\it potentialProto.h} file) and the field-space tensorial quantities into the {\it PyTransport/CppTrans/fieldmetric.h} file (using the {\it fieldmetricProto.h} file). A line in the {\it potential.h} file is updated every time this function is run to record the 
time of the last run.

\item {\it \bf PyTransSetup.compileName(Name,NC)} takes the argument ``Name" which must be a string and an optional boolean value ``NC". If the model is non-canonical the value of NC is {\color{blue}\bf{True}}, if not {\color{blue}\bf{False}} (by default set to false). This function complies the \CC \S  code into a Python module called ``{PyTransName}" using the file {\it PyTransport/PyTrans/moduleSetup.py} which it automatically edits. The module is placed in the folder {\it PyTransport/PyTrans/lib/Python/}. Note that this function also automatically makes edits to the {\it PyTrans/PyTrans.cpp} file. This function is intended to be used on Python 2 systems.

\item {\it \bf PyTransSetup.compileName3(Name,NC)} performs the same tasks as the function above, but 
is intended to be used on Python 3 systems.

\item {\it \bf PyTransSetup.deleteModule(Name)} takes the argument ``name" and deletes the compiled module {\it PyTransName}.
\end{itemize}

\section*{Appendix 3: Compiled module }

The functions which are part of the complied PyTrans module are detailed below. 

These functions essentially use the \CC \S  code in the {\it CppTrans/} folder to perform various tasks. 
The code for the functions can be found in the file {\it PyTrans/PyTrans.cpp} file. This code 
should be clear to those who are familiar with \CC. The only elements which might unfamiliar are those which 
convert the \CC \S  arrays to numpy arrays (at input and output), and the part of the file which defines how distutils 
is to turn the code into a Python module. A guide to creating compiled Python modules which we followed can be found \href{http://www.tutorialspoint.com/python/python_further_extensions.htm}{at this tutorial page}\footnote{\rm http://www.tutorialspoint.com/python/python\_further\_extensions.htm}.

One other point to note is that the code which defines the functions which evolve correlations 
invokes a scaling of the wavenumbers involved (the kscale variable in the code), 
which helps the integrator perform well for very different input values of the wave numbers. The rescaling is then reversed 
for output values which are returned to the python code calling the function.

Note the *** (Name) label needs to be added to the functions below if the module has been given a name, as described in the main text.

\begin{itemize}
\item    {\it \bf PyTrans.nF()} takes no arguments and returns an integer which is the number of fields present in the model.  
\item    {\it \bf PyTrans.nP()} takes no arguments and returns an integer which is the number of parameters used to define the model.
\item    {\it \bf PyTrans.V(fields, params)} takes a numpy array ``fields" of length nF (the number of fields) containing a set 
    of field values, and a numpy array ``params" of length nP (the number of parameters) containing parameter values. 
    It returns a double which is the value of the potential for these field values and parameters.
\item  {\it \bf PyTrans.dV(fields, params)} takes a numpy array of length nF (the number of fields) containing a set 
    of field values, and a numpy array of length nP (the number of parameters) containing parameter values. It returns a numpy array of length nF, which contains the first derivative of the potential.
\item  {\it \bf PyTrans.dVV(fields, params)} takes a numpy array of length nF (the number of fields) containing a set 
    of field values, and a numpy array of length nP (the number of parameters) containing parameter values. It returns a two dimensional 
    numpy array of size nF by nF, which   
    contains the second derivatives of the potential.
\item  {\it \bf PyTrans.H(fields-dotfields, params)} takes an numpy array ``fields-dotfields" of length 2 nF (twice the number of 
    fields) containing a set of field values followed by the field's velocities in cosmic time 
    (field derivative wrt cosmic time), and an array of length nP  containing parameter values.
    It returns as a double the value of the Hubble rate.
\item  {\it \bf PyTrans.backEvolve(Narray, fields-dotfields, params, tols, exit)} takes a numpy array of times in efolds ``Narray"
    at which we want to know the background value of the fields and field velocities. 
    This must start with the initial time in e-folds (initial N) we wish to evolve the system from, and finish with the 
    final value of N. It takes  a numpy array of length 2 nF ``fields-dotfields" which contains 
    the background field and velocity values at the initial time, as well as 
    a numpy array of length nP  ``params" containing parameter values.
It also takes an array ``tols" containing the relative and absolute tolerance and a boolean which if {\color{blue}\bf{True}} will exit the numerical evolution when inflation ends ($\epsilon =1$) or if {\color{blue}\bf{False}} continues until the desired number of e-folds has elapsed.
      It returns a two dimensional numpy array. 
    This array contains the fields, and field velocities at the times contained in  
    Narray. The format is that the array has 1 + 2nF columns, with the zeroth column the 
    times (Narray), the next 
    columns are the field values and field velocity values at those times. 
\item  {\it \bf PyTrans.sigEvolve(Narray, k, fields-dotfields, params, tols, full)} takes a numpy array of times in efolds ``Narray" 
    at which we want to know the value of the two point function of inflationary perturbations. 
    This must start with the initial N we wish to evolve the system from, and finish with the 
    final N. It also takes a Fourier mode value 
    ``k", and the initial conditions of the background cosmology (field and field velocity values) at the initial time  
    as a numpy array of length 2 nF, the parameters of the system as a numpy array of length 2 nP, an array ``tols" containing the relative and absolute tolerance,
    and an integer ``full" set to {\color{blue}\bf{False}} or {\color{blue}\bf{True}}
    (if another value it defaults to {\color{blue}\bf{True}}). The initial time and the initial field and field velocity array 
    are used to calculate the initial conditions for the evolution of the two-point function. 
    The function returns a two dimensional numpy array 
     the zeroth column of which contains the times (Narray). If full={\color{blue}\bf{False}} the 
     next column contains the power spectrum 
     of $\zeta$ at these times, and this is the final column 
     (there are therefore only 2 columns in total).
     If full = {\color{blue}\bf{True}} then there is in addition 2 nF + 2 nF * 2 nF columns. The first 2 nF of these columns 
     contain 
     the fields and field velocities at the time steps requested. Then the final 2 nF * 2 nF
 contain  the  elements of matrix $\Sigma_r^{ab}$ (the power 
     spectrum and cross correlations of the fields and field velocities). There are therefore [1 + 1 +2 nF +2 nF *2 nF] columns in total.
  	The convention is that the element $\Sigma_r^{ab}$ corresponds to the [ 1 + 2nF  + a + 
    2nF x (b-1)]th column of the array (recall the columns start at the zeroth column -- a and b run from 1 to 2 nF). 

\item  {\it \bf PyTrans.alpEvolve(Narray, k1, k2, k3, fields-dotfields, params, tols, full)} takes a numpy array of times in efolds (N) 
    at which we want to know the value of the three point function of inflationary perturbations. 
    This must start with the initial N we wish to evolve the system from, and finish with the 
    final N. It also takes three Fourier mode values 
    (k1, k2, k3), the initial conditions of the background cosmology (fields and field velocities) at the initial time as a numpy array of length 2 nF, the parameters of the system as a numpy array of length 2 nP, an array ``tols" containing the relative and absolute tolerance, 
    and an integer ``full'' set to {\color{blue}\bf{False}} or {\color{blue}\bf{True}}
    (if another value it defaults to 1). 
    The function returns a two dimensional numpy array, 
     the first column of which contains the times (Narray). 
    If full={\color{blue}\bf{False}} the next four columns contains the power spectrum of zeta 
     for each of the three k values input, and the value of the 
     bispectrum of zeta for a triangle with sides of length of these k values. A total of [1+  4] columns. If full={\color{blue}\bf{True}}, there are 
     an additional 2*nF + 6 *2  nF* 2 nF + 2 nF * 2 nF * 2 nF columns. The first 2nF  of these columns contain 
     the fields, and field velocities at the time steps requested (the background cosmology). The next 2nF * 2nF of these contain the real 
    parts of $\Sigma^{ab}(k_1)$ in the same numbering convention as above. Then the real part of 
    $\Sigma^{ab}(k_2)$ and then the real parts of $\Sigma^{ab}(k_3)$, the following 3 * 2nF * 2nF columns are 
    the imaginary parts of the $\Sigma^{ab}(k_1)$, $\Sigma^{ab}(k_2)$ and $\Sigma^{ab}(k_3)$ matrices. So for 
    example if one wanted access to the $\Sigma_i^{ab}(k_2)$ element that would be the 
    [1 +4 + 2nF + 4 * 2nF * 2nF + a + 2nF * (b-1)]th column of the file. The final 
    2nF * 2nF * 2nF columns of this file correspond to the $B^{abc}(k_1,k_2,k_3)$ matrix, such that     
    the corresponding columns would be the [1 + 2nF + 6 * 2nF * 2nF + a + 2nF * (b-1) + 
    2nF * 2nF * 2nF * (c-1)]th columns (recall the columns start at the zeroth column -- a, b and c run from 1 to 2 nF). 

\end{itemize}

\section*{Appendix 4: Python Scripts}

The functions which are part of the PyTransScripts modules are detailed below. The code 
which provides these scripts in the {\it PyTransScripts.py} file should be clear to those familiar with Python. The scripts are  
simply an indication of what is possible, and it is intended that users will modify them for their own purposes, or 
write their own, as well as using the ones provided. 

\begin{itemize}
\item   { \it \bf PyTransScripts.ICsBE(NBExit, k, back, params, PyT)} takes in the number of e-folds before horizon exit of a scale k at 
which initial conditions for the evolution of correlations are to be set, the scale k itself, a numpy array ``back" containing 
the background cosmology (returned from {\it PyTrans.backEvolve}), the parameters of the model
as a numpy array, and the PyTrans module being used. It returns a double, and an numpy array. The former is 
the starting number of efolds which are at least NBExit before exit, as measured with respect to the background trajectory, ``back". The array contains the initial value of the fields and field velocities at this starting time. This script simply runs through the elements of 
back to find the first one before the exit time minus NBExit. As such 
it requires back to have enough entries for the result to be 
useful (roughly 10 per e-fold is fine), as discussed in the main text. 

\item { \it \bf PyTransScripts.ICsBM(NBMassless,k, back, params, PyT)} works like PyTransScripts.ICsBM, but instead of calculating initial conditions before 
exit time, it evaluates the time when $k^2/a^2 = M^2$ where $M$ is the largest eigenvalue of the mass matrix of the potential (we call this the massless condition), and returns conditions 
more than NBExit before that time. This is useful for example with a heavy field, for which we need to ensure the approximation we use to 
fix initial conditions is accurate (which requires $k^2/a^2 \gg M^2$).

\item { \it \bf PyTransScripts.ICsBM(NB,k, back, params, PyT)}  takes the same arguments as the two previous 
functions and calls each in turn, then outputs the number of e-folds and the fields and field velocities at the earliest time of the two. 
This is so that the start time can be set to either NB before exit or NB before the massless condition, whichever  is earlier.

\item { \it \bf PyTransScripts.kexitN(Nexit, back, params, PyT)} takes an exit time in e-folds, a background evolution which runs though this time (returned from {\it PyTrans.backEvolve}),  
a set of parameters and the PyTrans module being used. It  returns as a double the k mode that exited at the time Nexit. This routine uses NumPy spline 
routines to find the value of k at the exit time.

\item { \it \bf PyTransScripts.kexitPhi(PhiExit, n, back, params, PyT)} takes an exit value of one of the fields, and a number indicating one of the fields (in range from 1 to nF), a background evolution which runs through this field value,  
a set of parameters and the PyTrans module being used. It returns the k mode that exited at the that field value. This routine uses NumPy spline routines to find the right k.

\item { \it \bf PyTransScripts.pSpectra(kA, back, params, NB, tols, PyT)}: takes a numpy array specifying  a range of k ``kA", a background evolution (output from the backEvolve function), a set of parameters, a double indicating the number of e-folds (before massless or before exit whichever is earlier) of sub-horizon evolution, an array containing the relative and absolute tolerance, and the PyTrans module being used. It returns two numpy arrays. The first has the corresponding values of $P_\zeta$ (to the input array of k) at the end of the evolution, and the second the times taken to perform the integration for each element.

\item { \it \bf PyTransScripts.pSpecMpi(kA, back, params, NB, tols, PyT)}: does the same as the function above but spreads the calculation across as many process as are active using Mpi4Py. The script which contains this function should be called using the the command ``mpiexec -n N python Script.py", where N is the number of processes to be opened. The length of kA must be divisible by N. We recommend calling at least twice as many processes as cores are available so that cores that run processes which finish first don't simply remain idle.  The function returns the two numpy arrays to the process with rank 0. Empty arrays are returned to the other processes.

\item { \it \bf PyTransScripts.eqSpectra(kA, back, params, NB, tols, PyT)}: takes in the same information as the function above, and returns three arrays. The first has the corresponding values of $P_\zeta$, the second the corresponding values of $B_\zeta$ in the equilateral configuration ($k1=k2=k3=k$) at the end of the evolution, and the final the times taken to perform the integration for each element.

\item { \it \bf PyTransScripts.eqSpecMpi(kA, back, params, NB, tols, PyT)}: does the same as the function above but spreads the calculation across as many process as are active using Mpi4Py. The script which contains this function should be called using the the command ``mpiexec -n N python Script.py", where N is the number of processes to be opened. The length of kA must be divisible by N. We recommend calling at least twice as many processes as cores are available so that cores that run processes which finish first don't simply remain idle.  The function returns the three numpy arrays to the process with rank 0. Empty arrays are returned to the other processes.

\item { \it \bf PyTransScripts.alpBetSpectra(kt,alpha, beta, back, params, NB, nsnaps, tols, PyT)}: takes in a value of kt, and two numpy  arrays defining a range of values of alpha and beta, as well as a background evolution, set of parameters and a number of e-folds before horizon exit or the massless condition is met and an array containing the relative and absolute tolerance. It also takes an integer nsnaps, and the PyTrans module being used. nsnaps  tells the code at how many different times at which to provide output. The function returns six arrays. 
The first output array is a three dimensional numpy array containing $B_\zeta$ for k values corresponding to the values of alpha, beta and kt  input, 
and at nsnaps different times between the start and finish time. The second array is also three dimensional and corresponds 
to $P_{\zeta}(k_1)$ for these values and times, the next $P_{\zeta}(k_2)$ and the next $P_{\zeta}(k_3)$. If nsnaps is 0 or 1, the third dimension of these arrays is only 1 element long, and the values returned correspond to those at the end of the evolution. If it is greater than one the output is given at evenly spaced times until the end of the evolution. This allows us to see how a slice through bispectrum evolves with time if we wish. 
The fifth array is two dimensional 
and corresponds to the times taken to perform the integrations associated with every combination of alpha and beta. The last array is the times at which 
the nsnaps are taken.

\item { \it \bf PyTransScripts.alpBetMpi(kt,alpha, beta, back, params, NB, nsnaps, tols, PyT)}: does the same as the function above but spreads the calculation across as many process as are active using Mpi4Py. The script which contains this function should be called using the the command ``mpiexec -n N python Script.py", where N is the number of processes to be opened. The length of alpha must be divisible by N. We recommend calling at least twice as many processes as cores are available so that cores that run processes which finish first don't simply remain idle.  The function returns the five numpy arrays to the process with rank 0. Empty arrays are returned to the other processes. 

\end{itemize}

\bibliography{paper}


\end{document}